\documentclass[12pt]{iopart}
\expandafter\let\csname equation*\endcsname\relax
\expandafter\let\csname endequation*\endcsname\relax
\usepackage{graphicx}
\usepackage{mathtools}
\usepackage{amsmath}
\usepackage{amssymb}
\usepackage{iopams} 
\usepackage{bm}
\usepackage{graphicx}
\usepackage{xcolor}
\usepackage{epstopdf}
\usepackage{dcolumn}
\usepackage{bm}
\usepackage{amsfonts,amssymb,amsmath,array}
\usepackage{amsbsy}
\usepackage{color}
\usepackage{hyperref} 
\usepackage{enumerate}
\usepackage{mathtools}
\usepackage{lipsum} 

\usepackage{bigints}
\usepackage{comment}
\usepackage{array}

\usepackage{soul}

\newcommand{\lx} {\left}
\newcommand{\rx} {\right}
\newcommand{\eps} {\epsilon}

\newcommand{\ave}[1] {\lx\langle #1 \rx\rangle}

\newcommand{\cP} {\mathcal{P}}
\newcommand{\cW} {\mathcal{W}}
\newcommand{\cS} {\mathcal{S}}
\newcommand{\cC} {\mathcal{C}}
\newcommand{\cG} {\mathcal{G}}
\newcommand{\cI} {\mathcal{I}}

\newcommand{\cK} {\mathcal{K}}
\newcommand{\cL} {\mathcal{L}}

\newcommand{\cF} {\mathcal{F}}

\newcommand{\cM} {\mathcal{M}}
\newcommand{\cN} {\mathcal{N}}

\newcommand{\cT} {\mathcal{T}}
\newcommand{\cQ} {\mathcal{Q}}

\begin{document}
\title[Statistical Features of Systems driven by non-Gaussian Processes]{Statistical Features of Systems driven by non-Gaussian Processes: Theory \& Practice}
\author{Dario Lucente$^1,^2$, Andrea Puglisi$^1,^2$, Massimiliano Viale$^1,^2$, Angelo Vulpiani$^1$} 

\address{$^1$ Dipartimento di Fisica - Universit\`a La Sapienza - 00185 Rome, Italy}
\address{$^2$ Istituto dei Sistemi Complessi - Consiglio Nazionale delle Ricerche, 00185 Rome, Italy}
\begin{abstract}
    Nowadays many tools, e.g. fluctuation relations, are available to characterize the statistical properties of non-equilibrium systems. However, most of these tools rely on the assumption that the driving noise is normally distributed. Here we consider a class of Markov processes described by Langevin equations driven by a mixture of Gaussian and Poissonian noises, focusing on their non-equilibrium properties. In particular, we prove that detailed balance does not hold even when correlation functions  are symmetric under time reversal. In such cases, a breakdown of the time reversal symmetry can be highlighted by considering higher order correlation functions. Furthermore, the entropy production may be different from zero even for vanishing currents. We provide analytical expressions for the average entropy production rate in several cases. We also introduce a scale dependent estimate for entropy production, suitable for inference from experimental signals. The empirical entropy production allows us to discuss the role of spatial and temporal resolutions in characterizing non-equilibrium features. Finally, we revisit the Brownian gyrator introducing an additional Poissonian noise showing that it behaves as a two dimensional linear ratchet. It has also the property that when Onsager relations are satisfied its entropy production is  positive although it is minimal. We conclude discussing estimates of entropy production for partially accessible systems, comparing our results with the lower bound provided by the \textcolor{black}{thermodynamic uncertainty relations}. 
\end{abstract}
\section{Introduction}
A deep comprehension of non-equilibrium systems
is one of the most relevant open  problems in statistical mechanics~\cite{KTH91}.
A crucial aspect of the \textcolor{black}{non-equilibrium} condition is the presence of currents induced 
by  some external constraints: physical currents - in the framework of Markov processes - imply that the detailed balance does not hold and, in general, that the time-reversal symmetry is statistically broken, or, 
equivalently, that the entropy production  is positive~\cite{BPRV08,Sekimoto2010,seifertrev,pelipigo}.
From a mathematical  point of view  the above statement can be assumed as   satisfactory
and it   has been thoroughly considered  in the context of several  Markov processes 
(e.g., Langevin equations and master equations), in particular an explicit expression for the entropy production
can be  introduced as the log-ratio between the probability of a long trajectory and that  of its time reversal~\cite{lebowitz1999gallavotti}.\\
Among the interesting issues which deserve investigation, one should include
the design of efficient  methods to characterise the  "degree of irreversibility", something also called "distance from equilibrium"~\cite{fodor2016far},
which typically requires a  proper modelling with a suitable mathematical  description of the system~\cite{seifert2019stochastic,skinner21,van2022thermodynamic,lucente2022inference,harunari2022learn,van2023time}. In this paper we discuss the role of Gaussian or non-Gaussian statistics for discriminating the degree of irreversibility of a system.\\
Starting from  the archetypal  Brownian Motion,
a plethora of phenomena have been  modelled and investigated  in terms of stochastic differential  equations with a Gaussian  random force (e.g. Gaussian white noise)~\cite{R89}. 
It is a matter of fact that Gaussian processes are ubiquitous in science, the reason is -  basically -  the central limit theorem, which can be succinctly summarised saying that a linear combination of many independent variables tends to behave as a Gaussian variable.
This - already at a qualitative level - is  a  strong argument for modelling the random forces appearing in stochastic differential equations in the form of Gaussian white noises.
For the same reason,  a large part  of  stochastic thermodynamics is devoted to models with such a kind of noise, which has been  successfully  adopted  even  for systems which are inherently out of equilibrium~\cite{Sekimoto2010,seifertrev,pelipigo}. There are several cases, for instance coming from the physics of driven granular gases or self-propelled particles,  where the usual linear Langevin equation is considered to be a satisfying approximation for the description of the dynamics of a massive probe, particularly in its  diffusive regimes~\cite{scalliet2015cages,lasanta2015itinerant,baldovin2019langevin,plati2020slow}.\\
More accurate analyses have shown, however, that - in some cases -  linear differential equations with \textcolor{black}{Gaussian white noises} 
are not able to describe some important features of the underlying dynamics, in particular they cannot catch the \textcolor{black}{non-equilibrium} statistical properties of the system. In order to restore 
non-equilibrium in the model, a suitable non-Gaussian noise is necessary~\cite{lucente2023revealing}.\\
At a first glance, the use of non-Gaussian noise can sound rather odd,
on the contrary  there are both   physical and  mathematical justifications for it.
For instance  we can consider a massive intruder  kicked by instantaneous collisions with agitated granular particles:
if the number of collisions in a given $\Delta t$ is not  very large,
the use of a Gaussian white noise is questionable. There are cases where it is more appropriate to take, as random force,
 a compound Poisson noise  $\zeta(t)$, see Section~\ref{sec:ts} for details.
In addition, it is possible to  show that non-Gaussian white noises like $\zeta(t)$  can be derived from microscopic theories through a systematic expansion of the Boltzmann-Lorentz equation governing the evolution of the intruder in a granular gas~\cite{kanazawa2012stochastic,kanazawa2015minimal,kanazawa2015asymptotic,kanazawa2017statistical}.
 Finally, we mention that,
  even from a mathematical point of view, by virtue of the Levy-Ito decomposition theorem, 
 the compound Poisson noise is one of the three contributions to Levy processes, i.e. processes with
  independent and identically distributed increments ~\cite{ken1999levy,kyprianou2014fluctuations,schilling2016introduction}. 
  \\
  The structure of the paper is the following.
  Section 2 is devoted to an analytical  and numerical investigation of systems with \textcolor{black}{non-Gaussian} forcing.
  We show  that it is possible to have a \textcolor{black}{non-equilibrium} system  even with  $\ave{x_i(t)\, x_j (0)}=\ave{x_j(t)\, x_i (0)}$:
  the breaking of the time \textcolor{black}{reversal} symmetry can  appear only looking at  other correlation functions e.g.  $\ave{x^3(t)\, x (0)}\neq \ave{x^3(0)\, x (t) }$.
  In a similar way the absence of a current is not sufficient for the time \textcolor{black}{reversal} symmetry to hold.
In Section 3  we study the entropy production $\cS$: it is possible to find an explicit expression
for systems driven by a forcing containing a  Gaussian term, with "temperature" $T$, and a compound Poisson noise.
It is interesting that in the absence of  a Gaussian noise,  the entropy production is infinite,
as a straightforward consequence  of the discontinuous character of the Poisson noise and of the dissipative dynamics between jumps.
Detailed numerical studies  of $\cS(\epsilon, \Delta t)$, i.e. the entropy production  computed at  space resolution $\epsilon$ and  sampling time $\Delta t$,
show that the convergence to the asymptotic value is very slow and a gigantic amount of data is necessary,  
an observation which has an immediate and practical relevance for the treatment of experimental signals.
  As a case study,  in  Section 4  we treat a $2D$ linear system driven by \textcolor{black}{non-Gaussian} forcing,
  that is a generalization of the Brownian Gyrator, comparing  $\cS$ with the average current and the
  deviations from the symmetry under time-reversal of higher order correlations. In Section 5 we draw conclusions and suggest perspectives.
In  the Appendix  we present some mathematical details for the computation of $\cS$.

\section{Time \textcolor{black}{reversal} symmetry and non-Gaussian noise}
\label{sec:ts}
In this section we investigate the effect of \textcolor{black}{non-Gaussian} delta-correlated noise on the equilibrium properties of stochastic processes, focusing our attention mainly on the time reversal symmetry. We will show that this noise generally drives the system away from equilibrium conditions even when fluctuation relations hold and we discuss \textcolor{black}{a strategy} to infer and measure the  ”degree of irreversibility" of the system.\\
Time reversal symmetry and thermodynamic equilibrium properties of a system are two strictly related concepts. 
In the framework of Markov processes such a relation is provided by the detailed balance condition~\cite{G90}. Indeed, a system described by a Markov process $X$ is said to be at equilibrium if 
\begin{equation}
    \pi(X)\cW_t(Y|X)=\pi(Y){\cW}_t(X|Y)
\label{eq:detail_balance}
\end{equation}
where $\cW_t(Y|X)$ is the probability to have $Y$ at time $t$ given the initial condition $X$ and $\pi(X)$ is the stationary probability. Condition~\ref{eq:detail_balance} implies that forward (in time) and backward paths have the same probability. Moreover, for any two functions $f,g$ that represent  (under time-reversal) even observables, one has $\ave{g(t)f (0) }=\ave{f(t) g (0) }$. When the system evolves according to a stochastic differential equation driven by Gaussian noise detailed balance also imposes vanishing currents. If one also restricts the class of investigated systems to Gaussian Markovian systems (for instance, in the continuous case, Langevin equations with linear forces and Gaussian noise) we have that the following conditions are sufficient for equilibrium~\textcolor{black}{\cite{Sekimoto2010,seifertrev}}:
\begin{itemize}
  \item  zero entropy production;
  \item no currents;
  \item time-reversal symmetry  of the correlation functions  i.e. $\ave{x_i(t)x_j(0)}=\ave{x_j(t)x_i(0)}$.
\end{itemize}
Actually  the above statements  are  equivalent;
it is quite natural to wonder about the effects of \textcolor{black}{non-Gaussian} forcing on the above scenario. \\
In the following we focus on the effect of a compound Poisson noise  $\zeta(t)$ on currents and time-reversal symmetry, postponing  the study of entropy production to the next section.
A compound Poisson noise $\zeta(t)$ is a stochastic process obtained as $\zeta(t) = \sum_j U_j \delta(t-t_j)$ where independent jumps, of random amplitudes ${U_j}$ ($U_j$ is a vector with the same dimensions of $X$) distributed according to $\mathcal{P}(U_j)$, occur at random times ${t_j}$. 
The differences $t_j-t_{j-1}$ are distributed according to $\cQ_\lambda(t)=\lambda e^{-\lambda t}$.
Such a noise arises naturally in granular system~\cite{kanazawa2012stochastic,kanazawa2015minimal,kanazawa2015asymptotic,kanazawa2017statistical,lucente2023revealing} and can also be used to model active forces~\cite{zakine2017stochastic,fodor2018non,bialas2023mechanism}. Moreover, similar noises have already been employed successfully both for modeling systems showing anomalous diffusion or stationary distribution with exponential tails~\cite{eliazar2003levy,barkai2020packets} or for implementing efficient protocols for finding "shortcuts to adiabadicity"~\cite{baldovin2022shortcuts}. 
Note that the properties of $\zeta(t)$ are strictly related to those of $\cP(U)$. In particular, if $\cP(U)$ has finite second moments, elements of $\Gamma=\ave{UU^T}$, then the central limit theorem holds and thus the sum of a large collection of jumps $\{U_j\}_{j\le N}$ tends to be normally distributed for large $N$, i.e.
\begin{align}
&\cP_N(z_N=z)\to\cG_\Gamma(z) = \frac{e^{-\frac{1}{2}z^T\Gamma^{-1}z}}{\sqrt{|2\pi\Gamma|}} \text{ as } N\to \infty\\
&z_N=\frac{1}{\sqrt{N}}\sum_{j=1}^N (U_j-\ave{U_j}).
\end{align}
Let us note that (assuming $\ave{U}=0$) correlations of $L(t)=\int_{0}^{t}\zeta(t')dt'$ are equivalent to those of a standard Wiener process~\textcolor{black}{\cite{rice1944mathematical,rice1945mathematical,hanggi1980langevin,van1983relation}}, i.e.
\begin{equation}
    \ave{L(t)L(t')} = \lambda \Gamma \,\,{\rm inf}\{t,t'\}.
\end{equation}
Such a result suggests that some properties of the system may not change when compound Poisson noise is used instead of the Gaussian one as a driving force. \\
Consider a stochastic process $X$ driven by a combination of a compound Poisson noise $\zeta$ and a Gaussian noise $\xi$, i.e.
\begin{equation}
    \dot{X} = F(X) + \xi(t) + \zeta(t) \qquad X=\{x_i\}_{i=1,N}
    \label{eq:Poisson_sde}
\end{equation}
with $\ave{\xi}=\ave{U}=0$, $\ave{\xi(t)\xi^T(t')}=2D\delta(t-t')$ and $\ave{UU^T}=\Gamma$.
Since the process $X$ is discontinuous, the detailed balance condition has to be imposed separately on the jumps and on the continuous part~\cite{G90}. 
Regarding the discontinuous part, \textcolor{black}{equilibrium} condition (\ref{eq:detail_balance}) takes the form
\begin{equation}
    \pi(X)\cP(Y-X)=\pi(Y)\cP(X-Y)\,.
    \label{eq:detailed_balance_poisson}
\end{equation}
\textcolor{black}{This means that if in the steady state the distribution of $X$ is spatially non-uniform ($\pi(X)\neq \text{const.}$) and the jumps are symmetric ($\cP(U)=\cP(-U)$) Eq.~\ref{eq:detailed_balance_poisson} can not be satisfied and the system is necessarily out of equilibrium.}
\begin{figure}[ht!]
\centering
    \includegraphics[width=0.48\textwidth]{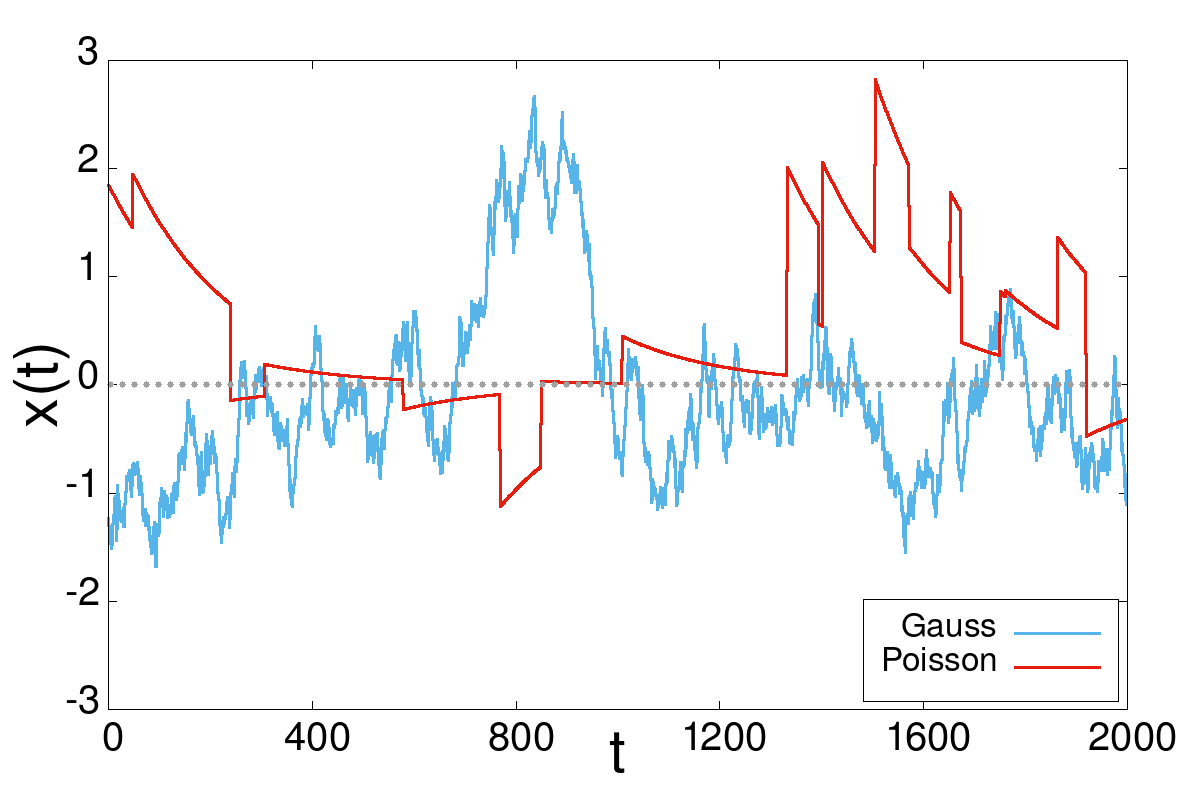}
    \includegraphics[width=0.48\textwidth]{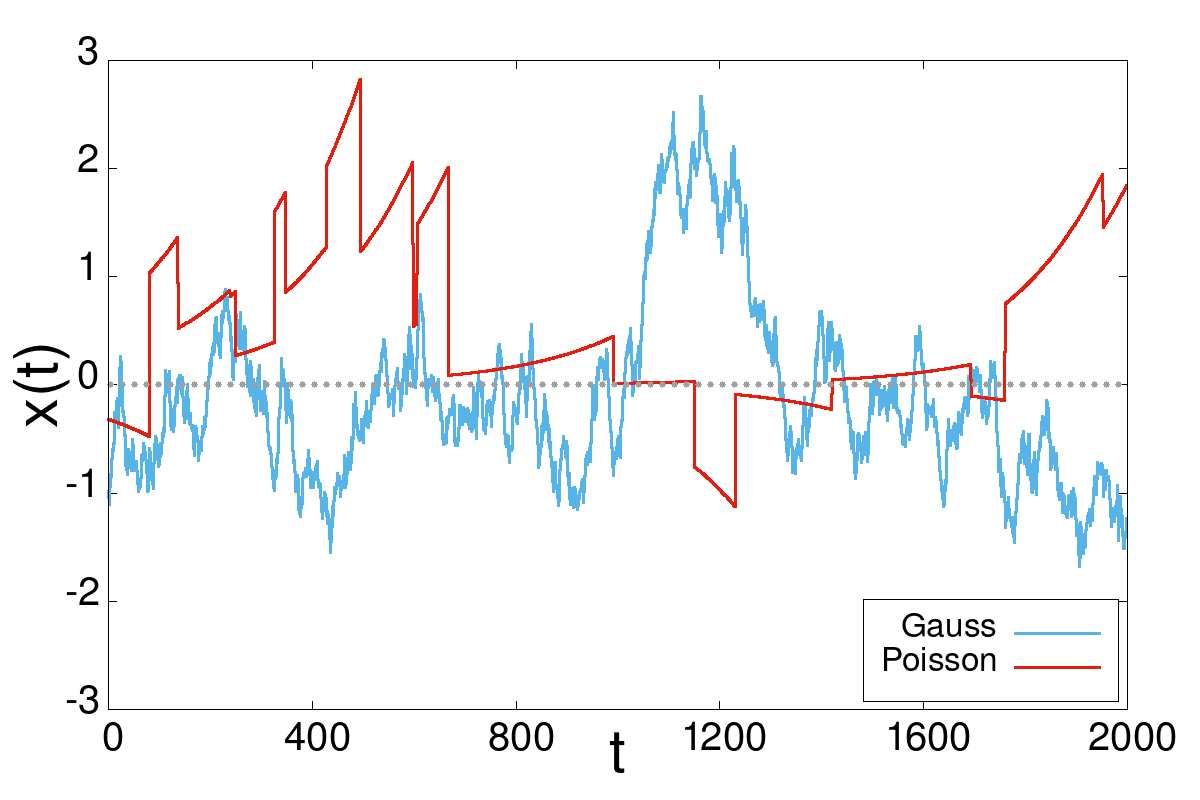}
  \caption{Examples of direct (left) and time-reverse (right) trajectories for processes driven by a Gaussian or Poisson noise. It is evident that in the Poisson case the time-reversed trajectory is strongly incompatible with the direct one, i.e. it is not possible to find any piece of the latter in the former. }
  \label{fig:Irreversibility_example}
\end{figure}
Nonetheless, if one restricts \textcolor{black}{one's} attention to linear systems ($F(X)=-AX$ with $A$ a positive definite matrix), $X$'s correlations  are equivalent to those of another system where the noises $\xi$ and $\zeta$ are replaced by a Gaussian white noise $\tilde{\xi}$ with $\ave{\tilde{\xi}(t)\tilde{\xi}(t')}=\lx(2D+\lambda \Gamma\rx)\delta(t-t')$. This means that, in this context, even if the Onsager relations ($\ave{x_i(t)x_j(0)}=\ave{x_j(t)x_i(0)}$) hold, they are no longer sufficient to determine whether that system is in thermodynamic equilibrium or not. Although this may seem surprising, recent works show that some thermodynamics relations (e.g. Einstein and Kubo) depend exclusively on the existence of a stationary state and therefore \textcolor{black}{hold} also in non-equilibrium conditions~\cite{yang2023time}. Statistical moments of order higher than two are necessary to discriminate between standard Gaussian noise and compound Poisson, and between equilibrium and non-equilibrium behavior~\cite{pomeau1982symetrie,lucente2023revealing}. Notwithstanding, sometimes the non-equilibrium nature of systems driven by compound Poisson noise can \textcolor{black}{be} easily grasped by looking at forward and backward time-series, as shown in Fig.~\ref{fig:Irreversibility_example}. Indeed, in the forward time-series (left panel) each jump is followed by a damping which instead appears as an acceleration in the backward time-series (right panel), signaling the time-reversal symmetry breaking.
In the following we will analyze in closer detail two one-dimensional examples, in order to  emphasize further the differences with respect to the equivalent Gaussian systems.

\subsection{1-D linear dynamics}
As a first example, let us consider a one-dimensional linear system
\begin{equation}
    \dot{X}=-\gamma X + \xi(t)+\zeta(t)
\end{equation}
with $\ave{\xi}=\ave{U}=0$, $\ave{\xi(t)\xi(t')}=2T\delta(t-t')$ and $\ave{U^2}=\Gamma$, with $\gamma$ and $\Gamma$ two positive parameters. \textcolor{black}{This equation has been proposed in~\cite{kanazawa2012stochastic, kanazawa2015minimal, kanazawa2015asymptotic} for describing the evolution of a massive probe in a granular medium and it has recently been shown~\cite{lucente2023revealing} that it can be thought as a minimal effective model for such a systems. Concerning this equation,}
we note that in the absence of $\zeta$ the system is necessarily at equilibrium (like all 1d systems without periodic boundary conditions) since it can not sustain any stationary current. In this case, the system is Gaussian and the first two moments fully characterize the statistics of $X$. The correlation function $\cC(t)=\ave{X(t)X(0)}$ is 
\begin{equation}
\cC(t)=\frac{T}{\gamma} e^{-\gamma t}\,.
\end{equation}\\
When the compound Poisson noise $\zeta$ is taken into account, the system is driven out of equilibrium and $X$ is no longer Gaussian. Nonetheless, given the confining nature of the potential, the system still can not sustain any net physical current  transporting the position steadily in a given  direction. Furthermore, the correlation function is symmetric \textcolor{black}{by} construction and takes the form
\begin{equation}
\cC(t)=\frac{2T+\lambda\Gamma}{2\gamma} e^{-\gamma t}
\end{equation}
which is exactly of the same form of the Gaussian case. On the other hand, non-trivial moments can display a breaking of time-reversal symmetry e.g. the $4-$th order correlation function $H(t)=\ave{X(t)X^3(0)}$. Indeed, for $t>0$ we have 
\begin{align}
&H(t)=\ave{X(t)X^3(0)}=e^{-\gamma t}\ave{X^4}\\
&H(-t)=\ave{X^3(t)X(0)}=e^{-3\gamma t}\ave{X^4}+3e^{-\gamma t}\lx(1-e^{-2\gamma t}\rx)\ave{X^2}^2
\end{align}
which are clearly different. \textcolor{black}{Note that $\langle X^4 \rangle$ ($\langle X^2 \rangle$) denotes the average of $X^4$ ($ X^2$) over the steady-state distribution $\pi(X)$.} Defining a degree of irreversibility $\Delta H (t)=H(t)-H(-t)$ leads to 
\begin{equation}
    \Delta H (t) = \lx(\ave{X^4}-3\ave{X^2}^2\rx)\lx(e^{-\gamma t}-e^{-3\gamma t}\rx).
\end{equation}
Note that the last equation correctly predicts $\Delta H(t)=0$ when the system is Gaussian.

\subsection{1-D dynamics on a ring}
\begin{figure}[ht!]
\centering
    \includegraphics[width=0.48\textwidth]{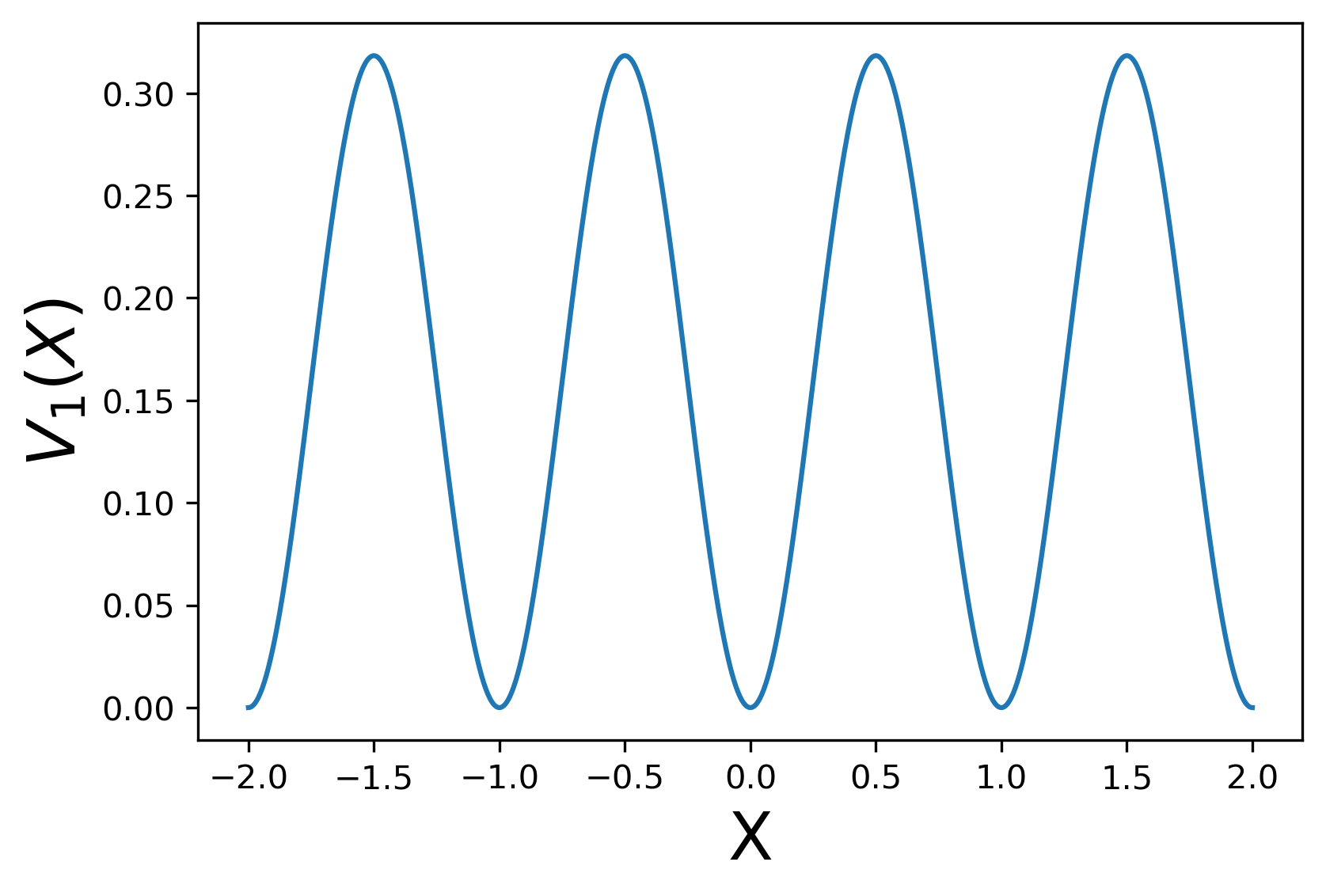}
    \includegraphics[width=0.48\textwidth]{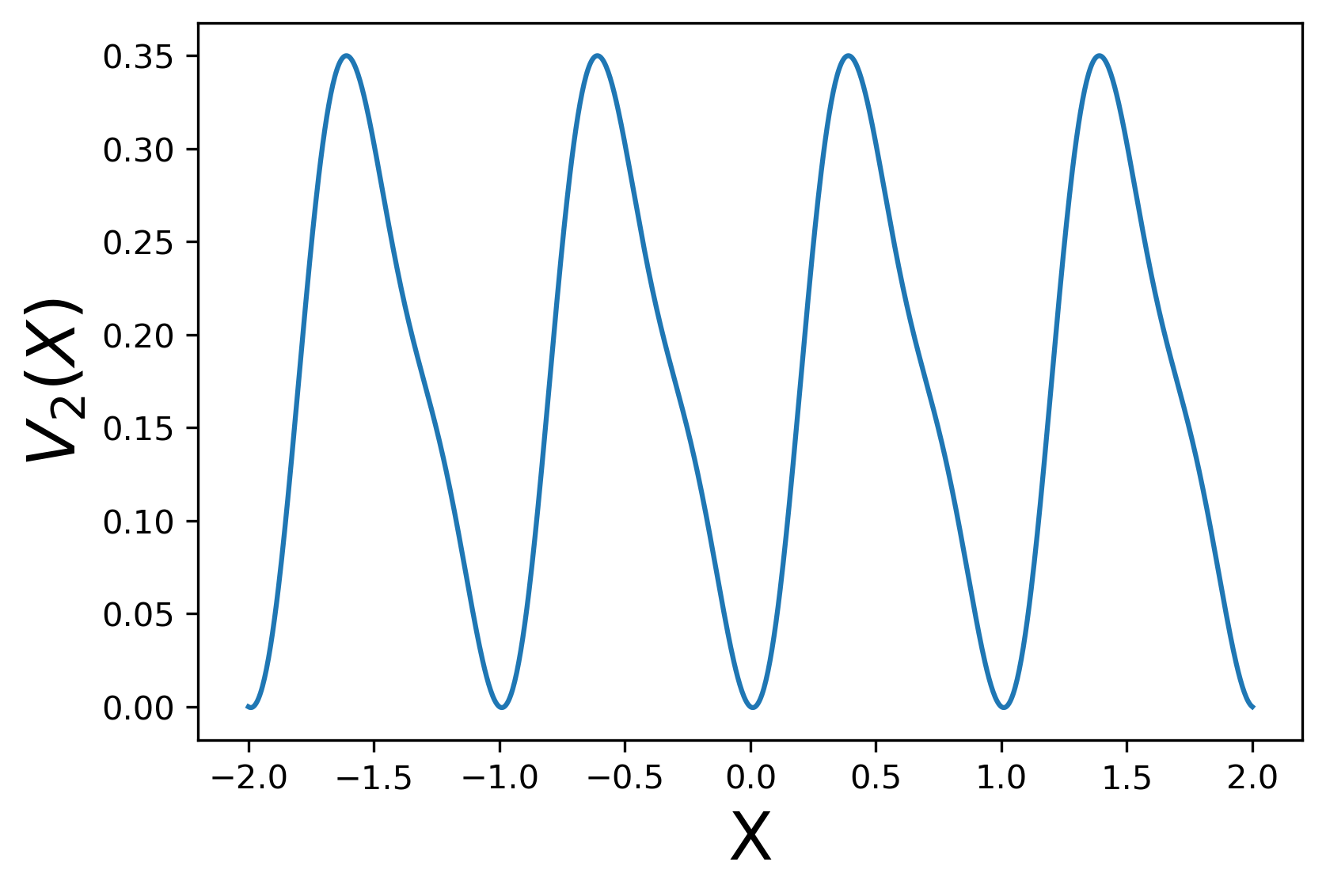}
  \caption{Potentials of Eqs.~\ref{eq:Potential}. Left: symmetric potential $V_1(X)$. Right: ratchet potential $V_2(X)$. 
  }
  \label{fig:Potential}
\end{figure}

\begin{figure}[ht!]
\centering
    \includegraphics[width=0.48\textwidth]{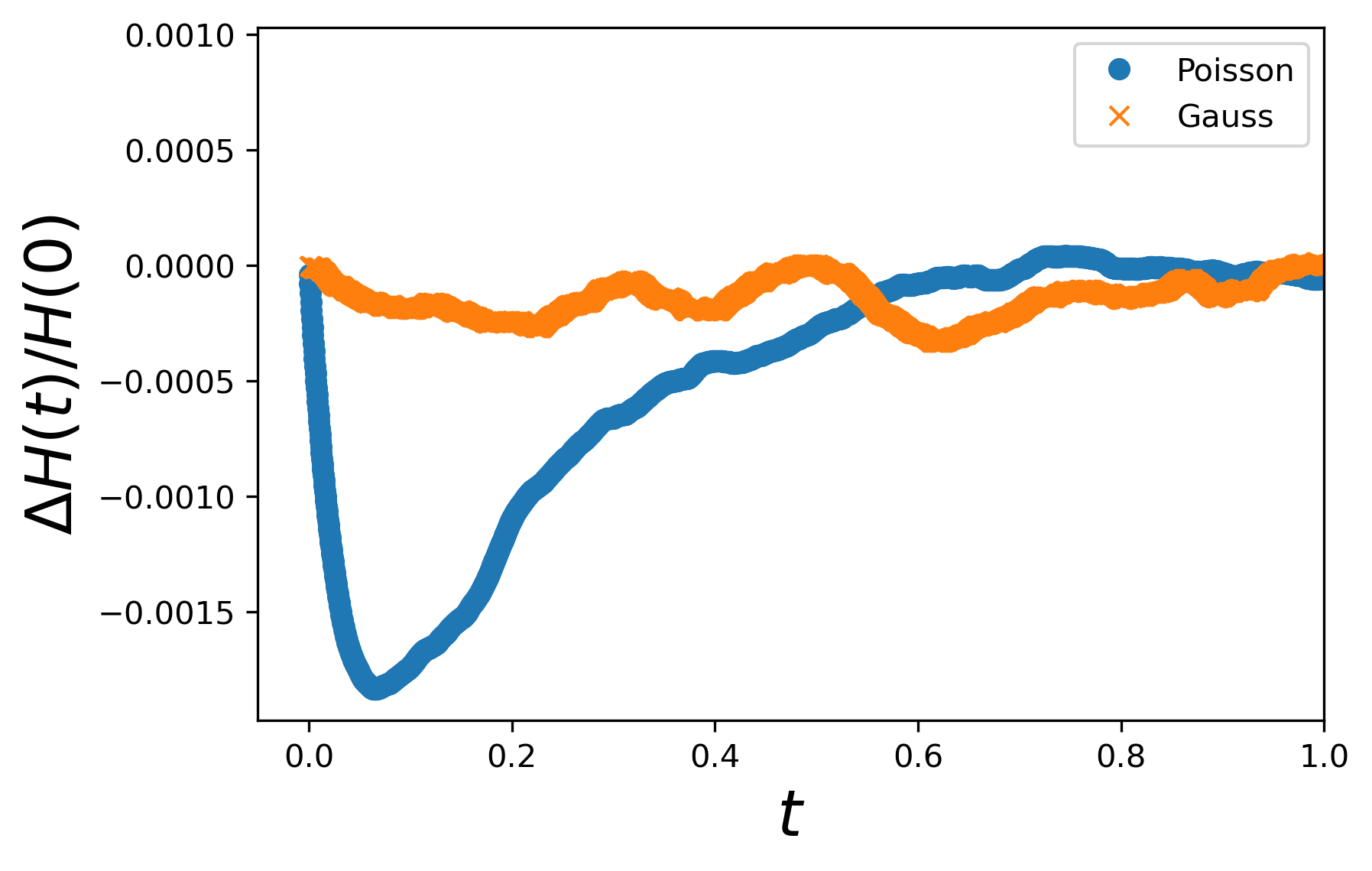}
    \includegraphics[width=0.48\textwidth]{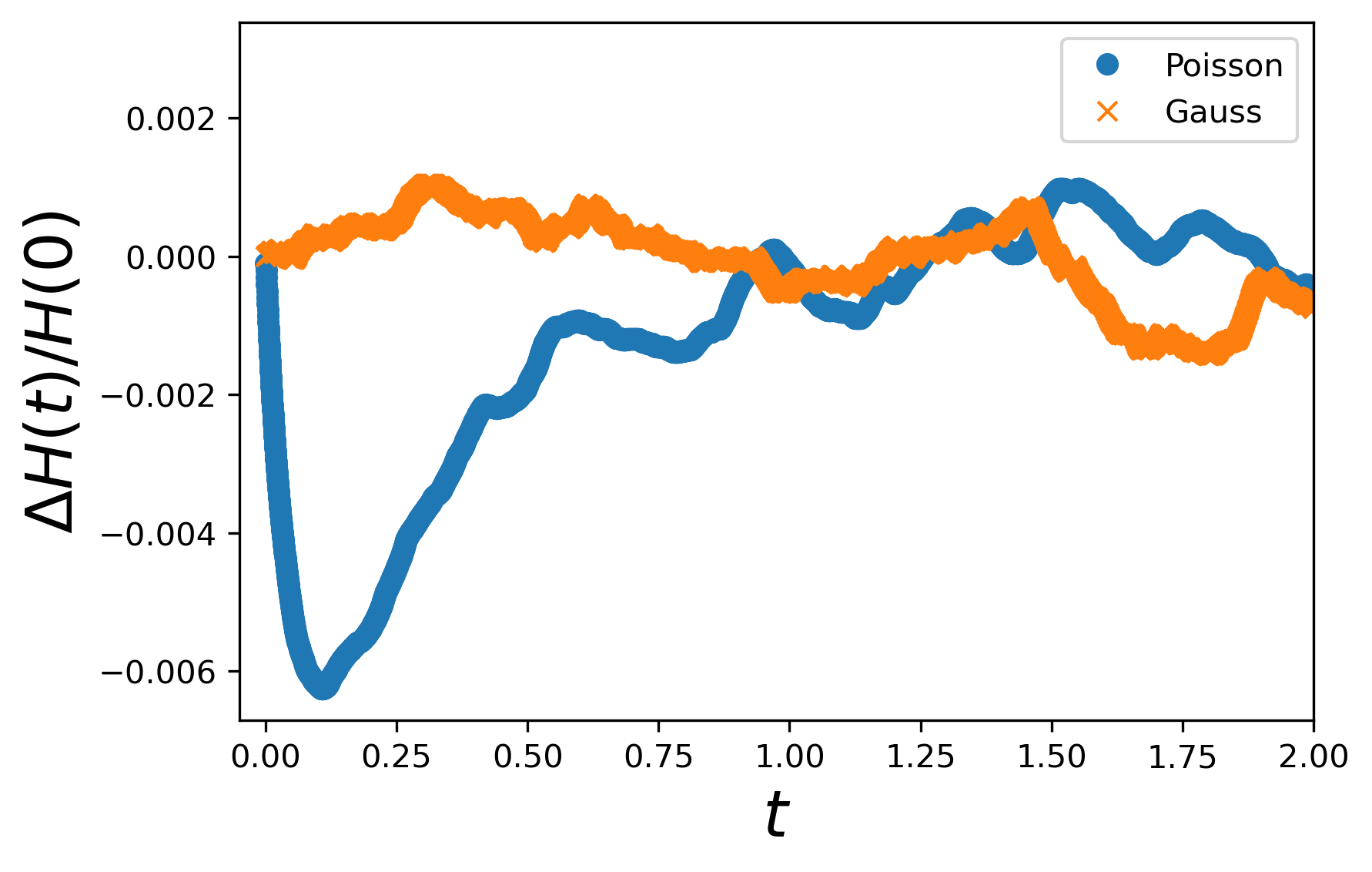}
  \caption{Degree of irreversibility $\Delta H(t)$ normalized to the value $H(0)$. The degree of irreversibility for the processes driven by compound Poisson noise $\zeta$ are shown in blue while orange curves represent the Gaussian cases. Left: symmetric potential. Right: ratchet potential. The parameters used for numerical simulations are $L=1,V_0=1,d=-0.2,\lambda=20,\sigma^2=0.2,T=0,\Gamma=\sigma^2/\lambda$. \textcolor{black}{The initial conditions are sampled from the stationary distribution.} 
  }
  \label{fig:Pomeau_1d_ring}
\end{figure}

\begin{figure}[ht!]
\centering
    \includegraphics[width=0.48\textwidth]{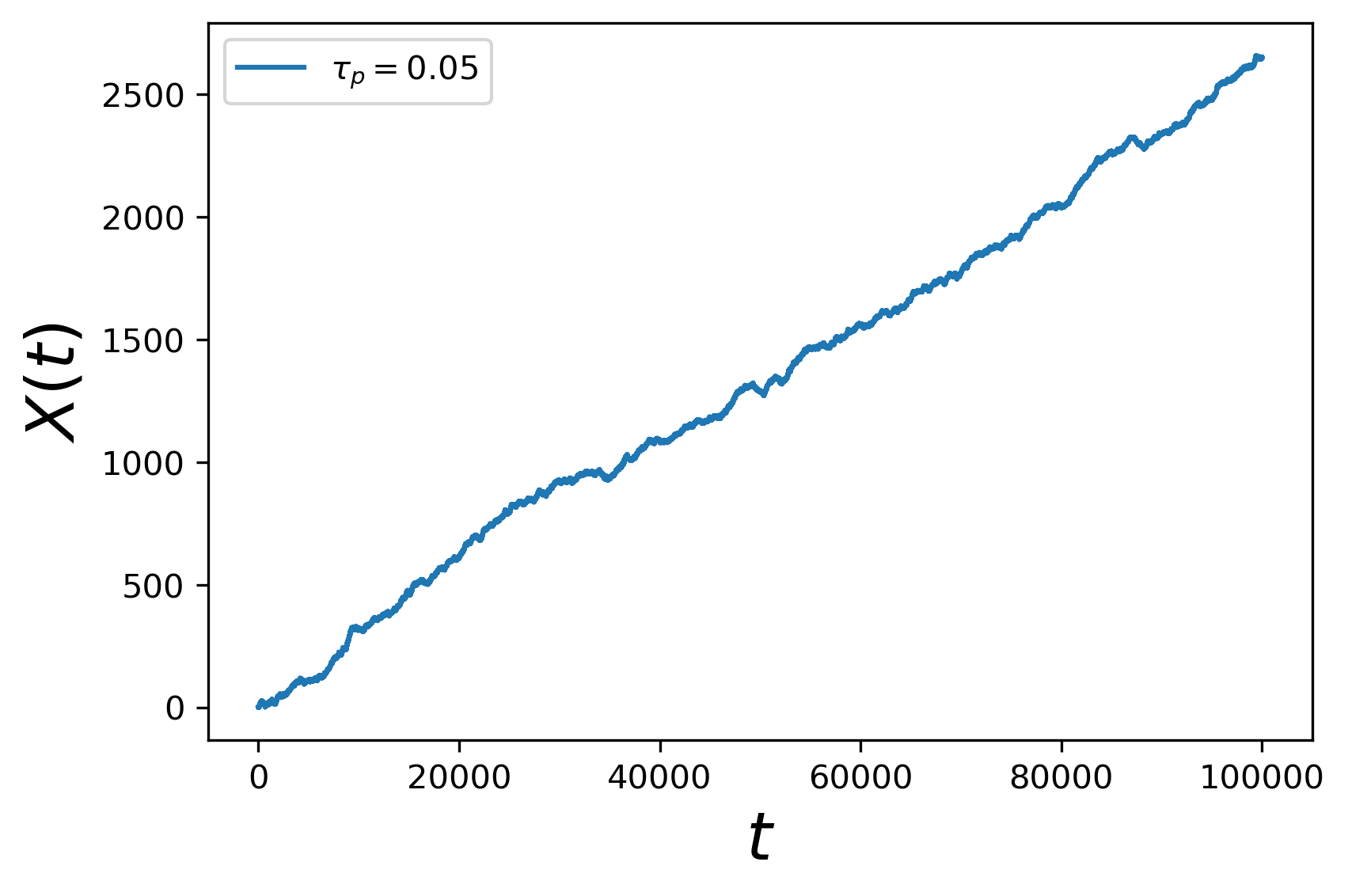}
  \caption{$X$ Vs $t$ for the process driven by the compound Poisson noise in the ratchet potential. The average slope represent the stationary current $j_s$. The parameters used for numerical simulations are $L=1,V_0=1,d=-0.2,\lambda=20,\sigma^2=0.2,T=0,\Gamma=\sigma^2/\lambda$. \textcolor{black}{The initial conditions are sampled from the stationary distribution.} 
  }
  \label{fig:Current_Ratchet}
\end{figure}

\textcolor{black}{One-dimensional} systems on a ring, e.g. with periodic boundary conditions, can sustain a non-equilibrium steady state and therefore constitute an excellent test-bed for investigating the effect of non-Gaussian noises on the properties of a system. \textcolor{black}{From a physical perspective, diffusion in periodic potential is used as a minimal model which displays transport phenomena and it might be relevant to discuss the role of external noise on these phenomena.}  Let us consider a system 
\begin{align}
    &\dot{X}=-\partial_X V(X) + \xi(t)+\zeta(t)\, ,\\
    & V(X+L)=V(X)\,.
\end{align}
This system has already been studied in \cite{luczka1997symmetric,kusmierz2018thermodynamics,bialas2020colossal,bialas2022periodic,bialas2023mechanism,spiechowicz2013absolute,spiechowicz2014brownian} and several non trivial behaviors arise. In particular, it has been shown that if the potential $V(X)$ has an asymmetric shape, the noise $\zeta$ induces a physical current in the system~\textcolor{black}{\cite{luczka1997symmetric}}. \textcolor{black}{Nonetheless, as a trivial consequence of the system laying in a \textcolor{black}{one-dimensional} space, the correlation function $\cC(t)$ is symmetric}.  Here we focus on the effects of $\zeta$ on the degree of irreversibility 
\begin{equation}
\Delta H(t)=\ave{X(t)X^3(0)}-\ave{X^3(t)X(0)}
\end{equation}
and its relation with the physical current. For this purpose, we consider two forms for the potential $V(X)$ (see Fig.~\ref{fig:Potential}), namely
\begin{align}
    & V_1(X)=\frac{L V_0}{2\pi}\lx(1-\cos{\frac{2\pi}{L}X}\rx) \,, \\
    & V_2(X)=\frac{L V_0}{2\pi}\lx(C+\sin{\frac{2\pi}{L}\lx(X-d\rx)}+\frac{1}{4}\sin{\frac{4\pi}{L}\lx(X-d\rx)}\rx)\,,\qquad C=\sin{\frac{2\pi d}{L}}+\frac{1}{4}\sin{\frac{4\pi d}{L}}.
    \label{eq:Potential}
\end{align}
where the parameter $d$ determines the minima positions. 
The first potential is symmetric and the noise $\zeta$ is not able to induce any current if the distribution of jumps $\cP(U)$ is an even function~\cite{luczka1997symmetric}. Notwithstanding this, $\Delta H(t)$ reveals the breaking of time-reversal symmetry in contrast to the Gaussian case where it is equal to zero, see the right panel of Fig.~\ref{fig:Pomeau_1d_ring}.\\
The potential $V_2(X)$ in Eq.~\ref{eq:Potential} instead has an asymmetric shape. Thus, the noise $\zeta$ induces a physical current as \textcolor{black}{is} clear in Fig.~\ref{fig:Current_Ratchet} that shows the variable $X$ drifting towards positive values. Clearly, the non-equilibrium nature of the system can also be revealed by the degree of irreversibility $\Delta H(t)$ (see Fig.~\ref{fig:Pomeau_1d_ring} right panel). These examples show once again that the absence of current and the symmetry of the usual (second order) correlation functions are necessary conditions for equilibrium, but are not sufficient if the system is driven by non-Gaussian noise.

\section{Entropy production and its empirical estimate}
\label{sec:ep}
In the previous section we have shown that stochastic differential equations driven by non-Gaussian white noises are out of equilibrium and the breaking of time reversal symmetry is revealed by looking at higher order correlation functions. This approach, satisfactory for the goal of simply discriminating, e.g. in experiments, between equilibrium and non-equilibrium, has some unavoidable disadvantages. In fact, although it suffices to find two functions $f,g$ for which one has \begin{equation}
\cC_{fg}(t)=\ave{f(0)g(t)}\neq \ave{f(t)g(0)}=\cC_{fg}(-t)
\end{equation}
in order to asses the non-equilibrium nature of the system, the degree of irreversibility $\Delta C_{fg}=\cC_{fg}(t)-\cC_{fg}(-t)$ depends on the functions $f$ and $g$ therefore it is not possible to introduce a quantity which is intrinsic i.e. that does not depend on the choice of the observables. This difficulty is overcome (at least from a formal point of view) by considering the entropy production $\cS$, which is an information-theoretic quantity. Formally, the entropy production $\cS$ of a Markov process $X$ is defined as~\textcolor{black}{\cite{lebowitz1999gallavotti}}
\begin{equation}
    \cS=\lim_{\cT\to\infty}\frac{1}{\cT}\ave{\log\lx(\frac{\cP(\{X_t\}_{0\le t \le \cT})}{\cP(\{ X_{\cT - t}\}_{0 \le t \le \cT})}\rx)}=\lim_{\cT\to\infty} \frac{\ave{\cS_\cT}}{\cT}
    \label{eq:epr}
\end{equation}
where $\cP(\{X_t\}_{0\le t \le \cT})$ represents the probability of the path $\{X_t\}_{0\le t \le \cT}$ in an analogous way $\cP(\{ X_{\cT - t}\}_{0 \le t \le \cT})$ of the reversed path $\{ X_{\cT - t}\}_{0 \le t \le \cT}$ and the average is done with respect to the forward path probability (we are always assuming, for simplicity, that all the relevant degrees of freedom are even under time-reversal). When the system admits a clear thermodynamics description, $\cS$ is related to the dissipated heat~\textcolor{black}{\cite{Sekimoto2010,seifertrev}}. For a discontinuous Markov process $X$, the probability of a given path $\{X_t\}_{0\le t \le \cT}$ is completely determined by the joint distribution of the discontinuities and the continuous paths between two subsequent discontinuities, that is (\textcolor{black}{see \ref{app:Setting} and \ref{app:Notation}})
\textcolor{black}{
\begin{equation}
\cP(\{X_t\}_{0\le t \le \cT})=\cP(\{X_t\}_{0\le t \le t_1^{-}})\cdots \cP(\{X_t\}_{t_n\le t \le \cT})\cP(t_1, \Delta X_{t_1}; \cdots,t_n,\Delta X_{t_n})
\label{eq:path_probability}
\end{equation}
where $\cP(t_1, \Delta X_{t_1}; \cdots,t_n,\Delta X_{t_n})$ represents the probability that discontinuous jumps $\Delta X_{t_k}=X_{t_k}-X_{t_k^-}$ for $k=1,\cdots,n$ occur at times $t_1,\cdots,t_n$. }
Thus, the quantity $\cS_\cT$ can be decomposed as
\begin{equation}
\cS_\cT = \cS_\cT^{c}+\cS_\cT^{j}
\label{eq:entropy_split}
\end{equation}
with
\begin{align}
&\cS_\cT^{c}=\sum_{i=0}^n \log\lx(\frac{\cP(\{X_t\}_{t_{i}\le t \le t_{i+1}^{-}})}{\cP(\{ X_{\cT -t}\}_{\cT-t_{i+1}^- \le t \le \cT-t_i})}\rx), \quad \quad t_0=0 \text{ and } t_{n+1}^{-}=\cT\,,\\
&\textcolor{black}{\cS_\cT^{j}=\log\lx(\frac{\cP(t_1, \Delta X_{t_1}; \cdots,t_n,\Delta X_{t_n})}{\cP(\cT-t_1, -\Delta X_{t_1}; \cdots,\cT-t_n,-\Delta X_{t_n})}\rx)\,,}
\end{align}
where $n$ is the number of discontinuities in the path. A similar decomposition has already been applied for computing the entropy production rate in a driven \textcolor{black}{one-dimensional} Lorentz gas~\cite{gradenigo2012nonequilibrium}. We note that decomposition (\ref{eq:entropy_split}) does not imply that the two quantities $\cS_\cT^{c}$ and $\cS_\cT^{j}$ are completely disjointed. 
In particular, the presence of discontinuities changes the probabilities for continuous paths so the statistics of $\cS_\cT^{c}$ also depends on the properties of \textcolor{black}{$\cP(t_1, \Delta X_{t_1}; \cdots,t_n,\Delta X_{t_n})$}. 
This result is valid for all Markov processes but often it is not very practical. There are a few cases for which it is possible to carry out analytical computations and obtain  simpler formula for $\cS$. Some stochastic process driven by a compound Poisson noise $\zeta$ belong to this class of Markov processes, as discussed in~\textcolor{black}{\ref{app:Linear-System} and \ref{app:Gradient-Systems}}.\\
\textcolor{black}{In our case, since the amplitude $U_k=\Delta X_{t_k}$ is independent from the time $t_k$ at which it occurs, the jumps' probability $\cP(t_1, \Delta X_{t_1}; \cdots,t_n,\Delta X_{t_n})$ takes the form
\begin{equation}
    \cP(t_1, \Delta X_{t_1}; \cdots,t_n,\Delta X_{t_n})=\cP(t_1,\cdots,t_n)\cP(U_{1})\cdots \cP(U_n)
\end{equation}
where $\cP(t_1,\cdots,t_n)$ is the uniform distribution over the region $t_1<t_2<\cdots<t_n<\cT$ (see \ref{app:Notation} for further details)}, which implies $\cS_\cT^{j}=0$ if $\cP(U)=\cP(-U)$. Thus, for stochastic processes driven by symmetric Poisson noise, one has $\cS_{\cT}=\cS^{c}_\cT$. In the following we will give the explicit expressions for the entropy production for the examples of the previous section. Before this, however, it is important, at least on a conceptual level, to discuss how this entropy production can be measured in real experiments. 
In the following we consider only symmetric $\cP(U)$. We will see that $\cS_\cT^c$ typically is determined by two contributions: one given by the Gaussian noise and the other by the Poissonian noise.

\subsection{Empirical estimate of the entropy production}
Estimating entropies from data is a difficult task, mainly due to the large number of data needed for having \textcolor{black}{a} precise result. The purpose of this section is to discuss both the technical and conceptual problems faced in entropy production measurements. Borrowing concepts from dynamical systems, we introduce the concept of scale-dependent entropy production $\cS(\epsilon,\Delta t)$ (\textcolor{black}{somewhat} analogous to $\epsilon$-entropy~\cite{abel2000exit,gaspard1993noise,boffetta2002predictability}) as follows:
\begin{itemize}
\item introduce a partition $\{C_i(\epsilon)\}_{1\le i \le K}$ of size $\epsilon$ of the phase space $\Omega_X$ (for example hypercube of side $\epsilon$),
\item define an empirical Markov chain on the space induced by the partition whose stationary probability $\pi_i$ and transition matrix $M_{ij}(\Delta t)$ are $$\pi_i=\cP(X_t\in C_i(\epsilon))\,,$$ $$M_{ij}(\Delta t)\textcolor{black}{= \cP(X_{t+\Delta t}\in C_j(\epsilon)|X_t\in C_i(\epsilon))} = \frac{\cP(X_t\in C_i(\epsilon),X_{t+\Delta t}\in C_j(\epsilon))}{\cP(X_t\in C_i(\epsilon))}\,,$$
\item compute the entropy production of this Markov chain $$\cS(\epsilon,\Delta t)=\frac{1}{\Delta t}\sum_{ij}\pi_iM_{ij}\log\lx(\frac{M_{ij}}{M_{ji}}\rx)\,.$$
\end{itemize}
Of course the Markov chain can be considered a good approximation only for $\Delta t$ and $\eps$ small enough.
The $\pi_i$ and the $M_{ij}(\Delta t)$ must be determined from a long trajectory.
Although the entropy production $\cS(\epsilon,\Delta t)$ is different from that of the real system $\cS$, in the limit $\epsilon \to 0$ and $\Delta t\to 0$ one has $\cS(\epsilon,\Delta t)\to\cS$~\footnote{For $\epsilon \to 0$ and $\Delta t\to 0$ the empirical stationary distribution $\pi$ and the transition matrix $M$ converge to their continuous counterparts, i.e. $\pi\to \pi(X)$, $M_{ij}\to \cW_t(Y|X)$. Thus, $\cS(\epsilon,\Delta t)\to \cS=\lim_{t\to 0} \frac{1}{t}\sum_{X,Y}\pi(X)\cW_t(Y|X)\log\lx(\frac{\cW_t(Y|X)}{\cW_t(X|Y)}\rx)$ which can be proven to be an equivalent definition for the entropy production of a Markov process~\textcolor{black}{\cite{Sekimoto2010,seifertrev,cocconi2020entropy} (see also \ref{appendix:EPR})}.}. \textcolor{black}{It is important to note that the $\cS(\eps,\Delta t)$ is not a lower bound for the entropy production. Indeed, in order to have a lower bound one should consider the entropy production of the non-Markovian coarse-grained process. Here, instead, we consider an approximate Markovian process and, as it is shown in~\cite{van2023time}, its entropy production could also be larger than the entropy production of the microscopic system. }
Let us now consider \textcolor{black}{in more detail} the expected behavior of $\cS(\epsilon,\Delta t)$ in the context of stochastic process driven by compound Poisson noise. The noise $\zeta$ has two relevant scales, one spatial and one temporal. The spatial scale is related to the typical size of jumps $U$, i.e. $\epsilon_{p}\sim\sqrt{\ave{U^2}}$, while the temporal one is dictated by the jumps rate $\lambda$, namely $\tau_p\sim\frac{1}{\lambda}$. In situations where $\Delta t \gg \tau_p$, one expects that $\cS(\epsilon,\Delta t)$ is in good agreement with the prediction of the Gaussian case, since there is time for a large number of jumps to occur and $\zeta$ is somehow close to a Gaussian noise. Similarly, for spatial resolutions greater than the size $\epsilon_p$ of typical fluctuations, Poisson noise \textcolor{black}{does} not contribute to the transitions between different cells of the partition. For small enough $\epsilon$ and $\Delta t$ instead, the quantity $\cS(\epsilon,\Delta t)$ is expected to be a good proxy of the continuous entropy production $\cS$. \textcolor{black}{All these observations are based exclusively on theoretical arguments and are therefore valid in the limit of an infinite amount of data. Problems can arise when the amount of data $N$ is limited \textcolor{black}{and so the results may not be statistically significant.} The most serious problem (and the only one we discuss briefly) is how to deal with missing transitions, i.e. situations for which $M_{ij}>0 $ but $M_{ji}=0$. In these situations, the definition of $\cS(\eps,\Delta t)$ leads to divergences. In order to avoid these divergences, a regularization can be applied, i.e. assuming a small but finite probability $\delta\ll\frac{1}{N}$ for the missing transitions. In this way, $\cS(\eps,\Delta t)$ takes only finite values that are almost independent on the value of $\delta$ (as long as $\delta\ll\frac{1}{N}$).}

\subsection{Entropy production for 1-D linear dynamics}\label{sec:epr_linear_1d}
\begin{figure}[ht!]
\centering
    \includegraphics[width=0.48\textwidth]{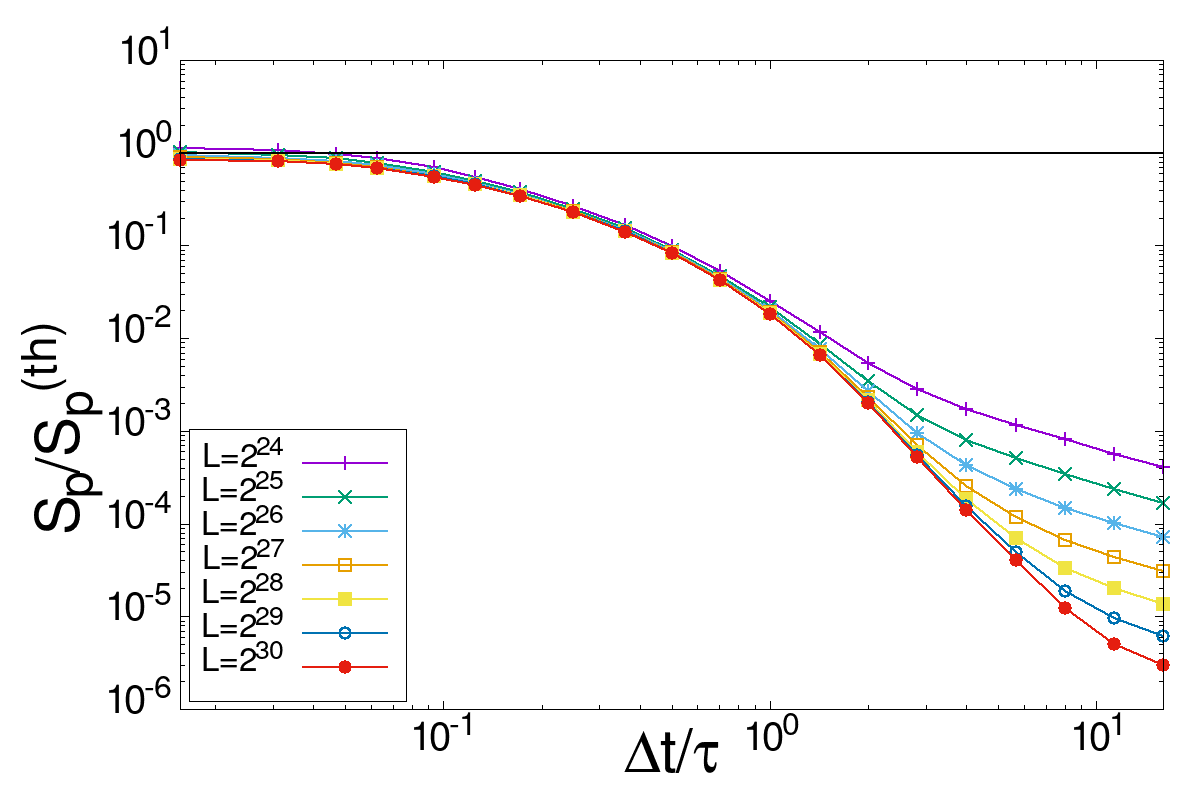}
    \includegraphics[width=0.48\textwidth]{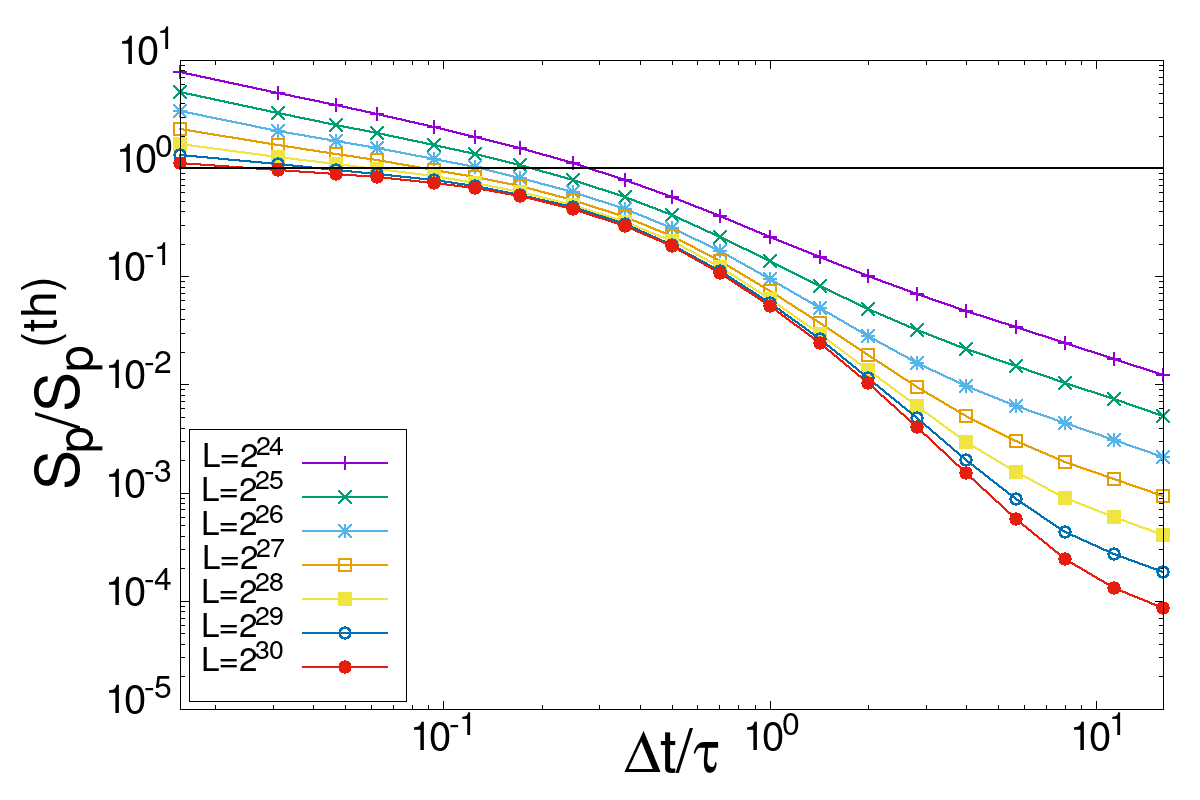}
 \caption{\textcolor{black}{Ratio between the empirical entropy production rate $\cS(\eps,\Delta t)$ and its theoretical value $\cS$ as a function of $\Delta t/\tau$} for $\eps=5.86\cdot 10^{-3}$. Different curves represent simulation of increasing duration. Horizontal black lines indicate \textcolor{black}{the asymptotic} value. 
  Left and right panels show two cases with different proportion between Gaussian and Poisson noise amplitude. 
  The parameters used for numerical simulations are
  $\gamma=1/180$, $\lambda=1/64$,  $(2T+\lambda\Gamma)=2\gamma$, $T=(1-p) \gamma$, $\Gamma=p (2 \gamma /\lambda)$, $p=0.99$ (left) $p=0.75$ (right). \textcolor{black}{The initial conditions are sampled from the stationary distribution.} }
  \label{fig:SpVsTau_Linear_Convergence}
\end{figure}

\begin{figure}[ht!]
\centering
    \includegraphics[width=0.48\textwidth]{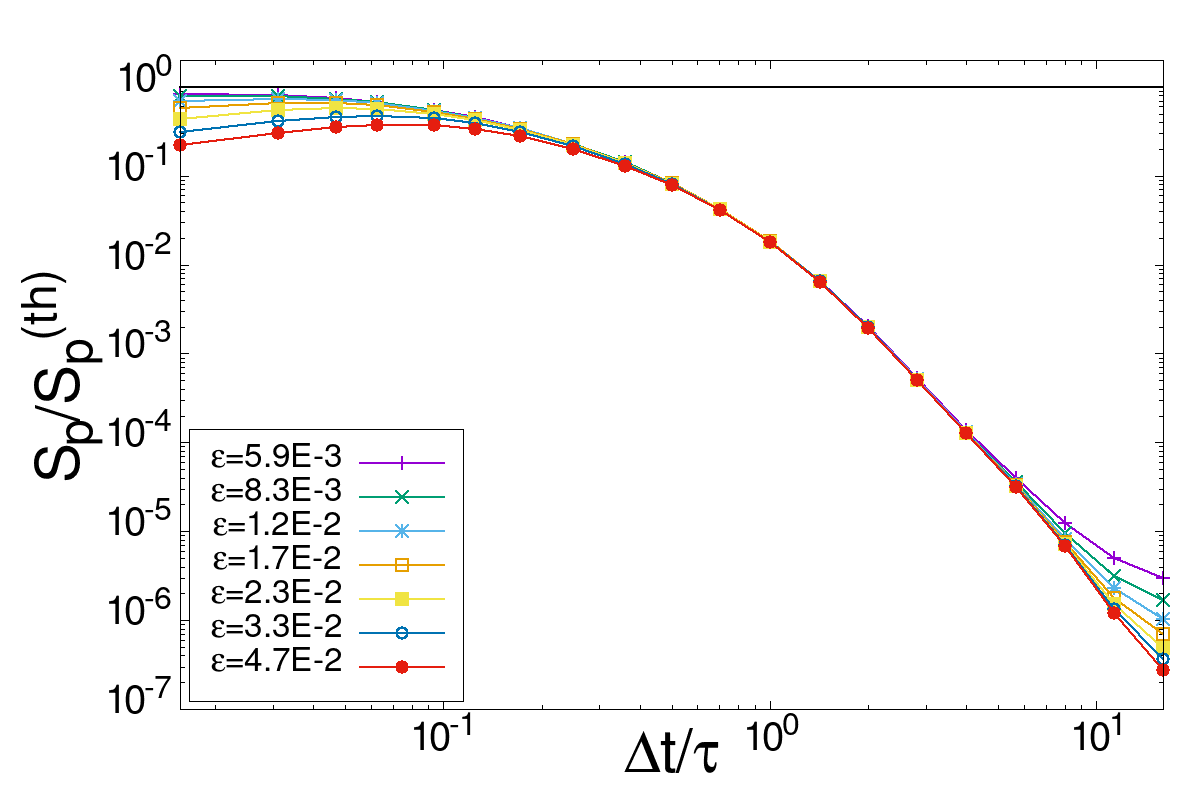}
    \includegraphics[width=0.48\textwidth]{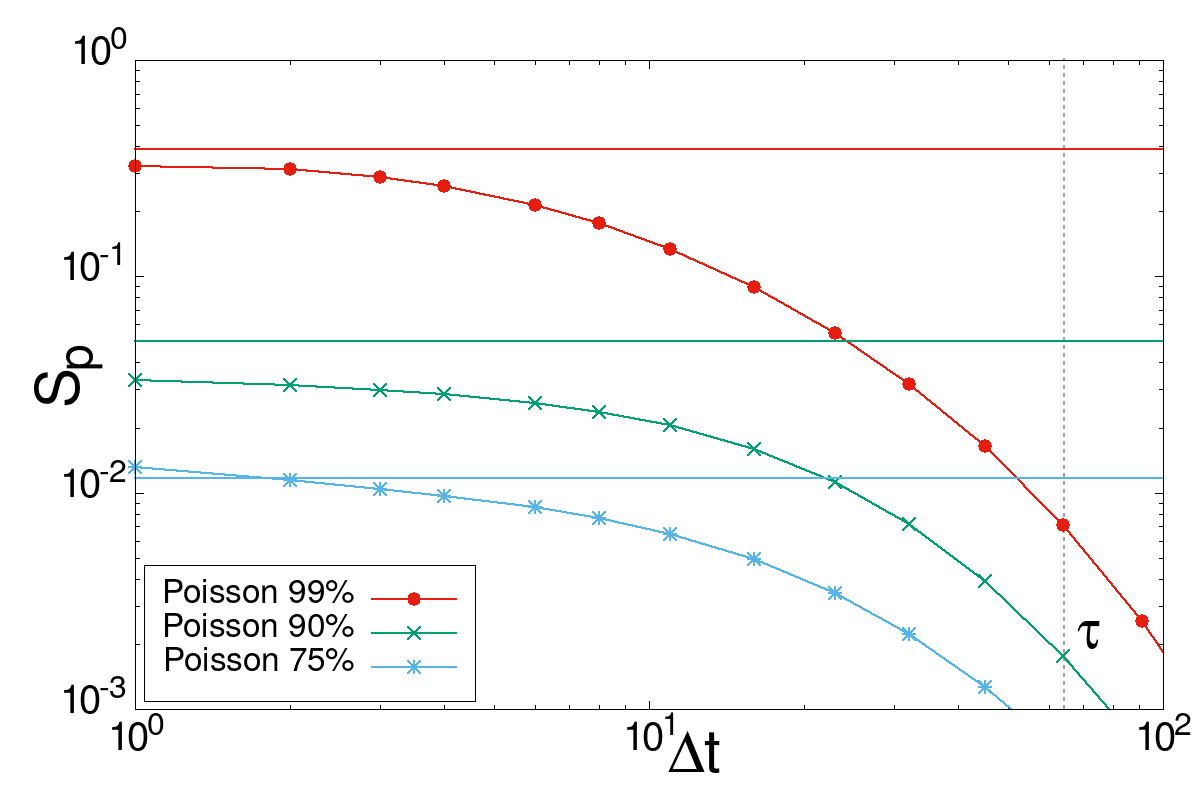}
  \caption{Left panel: \textcolor{black}{Ratio between the empirical entropy production rate $\cS(\eps,\Delta t)$ and its theoretical value $\cS$ as a function of $\Delta t/\tau$ for several $\eps$. Horizontal black line indicate the asymptotic value. Right panel: convergence of $\cS(\eps,\Delta t)$ towards the theoretical value (horizontal lines) for different level of Poisson noise, $p=0.75,0.90,0.95$. The vertical dashed black line represents the Poisson jump rate $\tau$}.}
  
  \label{fig:SpVsTau_Linear}
\end{figure}

Let us discuss the effect of non-Gaussian noise on entropy production in \textcolor{black}{one-dimensional} linear systems. Given the simplicity of linear models, this example allows us to understand the main difficulties encountered in estimating entropy production. 
Moreover, analytical computations for entropy production can be easily performed. \textcolor{black}{The details can be found in~\ref{app:Gradient-Systems} while the result is}:

\begin{align}
   \cS_\cT^{c}&=T^{-1}\lx[V(X_{t_{0}})-V(X_{\cT})\rx] +T^{-1}\sum_{i=1}^n\lx[V(X_{t_{i}}+U_i)-V(X_{t_i})\rx]\,\nonumber \\
   &=T^{-1}\lx[V(X_{t_{0}})-V(X_{\cT})\rx]+\frac{\gamma}{2T}\sum_{i=1}^n\lx[U_i^2+2X_{t_i}U_i\rx]
   \label{eq:entropy_linear}
\end{align}
In the limit $\cT \to \infty$ we have that, for $\cS_\cT^{c}/\cT$ the first term contains only boundary contributions and therefore it is zero, on the contrary, the second term of (\ref{eq:entropy_linear}) increases proportionally to $n$ which on average is equal to $\lambda \cT$. Thus, the entropy production rate $\cS$ (Eq.~\ref{eq:epr}) is
\begin{equation}
\cS=\frac{\gamma\lambda\Gamma}{2T}.
\end{equation}
At this point we can discuss the empirical estimate $\cS(\epsilon,\Delta t)$ and test whether it is able to give a reasonable proxy to the entropy production $\cS$. Figs.~\ref{fig:SpVsTau_Linear_Convergence} show the empirical entropy production $\cS(\epsilon,\Delta t)$ as a function of the temporal scale $\Delta t$ as the duration of the trajectories $\cT$ increases. 
\textcolor{black}{As one can see, when the true value of entropy production is not too "small" \textcolor{black}{(order $\sim 10^{-1}-1$, see the red curve on right panel of Fig.\ref{fig:SpVsTau_Linear})}, we are able to get a good numerical agreement, a relative error of the order $10\%$, even by computing $\cS(\epsilon,\Delta t)$ with short trajectories. Conversely, although the absolute error is about the same, if the true entropy production is too small \textcolor{black}{(order $\sim 10^{-2}$, see the light blue curve on right panel of Fig.\ref{fig:SpVsTau_Linear}))}, a gigantic amount of data is needed to have the same accuracy.} This can impose severe limitations in experiments with more complicated systems having multiple time scales. \textcolor{black}{Note, however, that this is just an empirical observation and different systems could behave differently one from each other. }
\textcolor{black}{}
\\
Left panel of Fig.~\ref{fig:SpVsTau_Linear} shows again $\cS(\epsilon,\Delta t)$ as a function of $\Delta t$ but for different value of $\epsilon$. As expected, differences between different \textcolor{black}{$\epsilon$} only appear when $\Delta t \ll\tau$ and the estimates become more and more accurate as $\epsilon \to 0$. Finally, right panel of Fig.~\ref{fig:SpVsTau_Linear} confirms once again that $\cS(\epsilon,\Delta t)$ is a good proxy in different situations and in particular when the system is "close to equilibrium" \textcolor{black}{($\cS\sim 10^{-2}$)}. 

\subsection{Entropy production for 1-D dynamics on a ring}
Consider now more complicated examples: the case where a particle is moving on a ring under the effect of a periodic potential and driven by both Gaussian and Poissonian noises. Here we employ the same potentials already used in Sec.~\ref{sec:ts} (Eqs.~\ref{eq:Potential}), with the difference that in the symmetric case we also add a constant pulling force $f$. Such a force breaks the time reversal symmetry even in the Gaussian case, inducing a stationary current $j_s$. It is well known that, in the Gaussian case, the entropy production $\cS$ is strictly related to the stationary current $j_s$ according to~\textcolor{black}{\cite{Sekimoto2010,seifertrev,cocconi2020entropy}}
\begin{equation}
\cS\propto \frac{j_s^2}{T}.
\end{equation}
Therefore, it seems natural to wonder how Poissonian noise changes this scenario. As already explained in the previous section, for gradient systems the change in entropy equals the change of internal energy. Since all \textcolor{black}{one-dimensional} system can be thought as gradient systems, the pulling force $f$ just \textcolor{black}{modifies} the internal energy as
$V(X)\to V_f(X)=V(X)-fX$.
The reader can find the details of the analytical computation in \textcolor{black}{\ref{app:Gradient-Systems}}.
The main result is the following 
\begin{equation}
\cS=\frac{j_s}{T}f + \Delta \cS_p\,.
\label{eq:relation_epr_current}
\end{equation}
In the case of symmetric potential we have
\begin{equation}
\Delta \cS_p = \frac{\lambda V_0 L}{2\pi T}\ave{\cos{\frac{2\pi}{L}x}} \lx(1-e^{-2\lx(\pi \Gamma/L\rx)^2}\rx)
\end{equation}
while in the case of the ratchet-like asymmetric potential, we have
\begin{equation}
\Delta \cS_p =-\frac{\lambda V_0 L}{2\pi T}\lx[\ave{\sin{\frac{2\pi}{L}(x-d)}} \lx(1-e^{-2\lx(\pi \Gamma/L\rx)^2}\rx)+\frac{1}{4}\ave{\sin{\frac{4\pi}{L}(x-d)}} \lx(1-e^{-8\lx(\pi \Gamma/L\rx)^2}\rx)\rx]\,,
\end{equation}
where $\ave{\cdot}$ indicates the average over the stationary distribution $\pi(x)$.
It is interesting to note that, independently of whether there is a physical current or not, the first term \textcolor{black}{of the right hand side of Eq.~\ref{eq:relation_epr_current}} vanishes when $f=0$. \\
Let us now consider the empirical entropy production $\cS(\epsilon,\Delta t)$, starting with the symmetric potential with a constant pulling force $f=0.5$. $\cS(\epsilon,\Delta t)$ \textcolor{black}{displays} several regimes depending on the timescales of relaxation and jump events. In particular, if the relaxation time $\tau_r\sim V_0^{-1}$ is much bigger than the jump rate $\tau_p=\lambda^{-1}$, we expect that $\cS(\epsilon,\Delta t)\sim j_s^2/T_{eff}$  for $\tau_p\ll\Delta t \ll \tau_r$ where $T_{eff}=T+\lambda\Gamma$. For $\tau_p\lesssim \tau_r$ the dynamics is dominated by the jumps and the empirical entropy production strongly \textcolor{black}{differs} from its Gaussian counterpart. 
\begin{figure}[ht!]
\centering
    \includegraphics[width=0.48\textwidth]{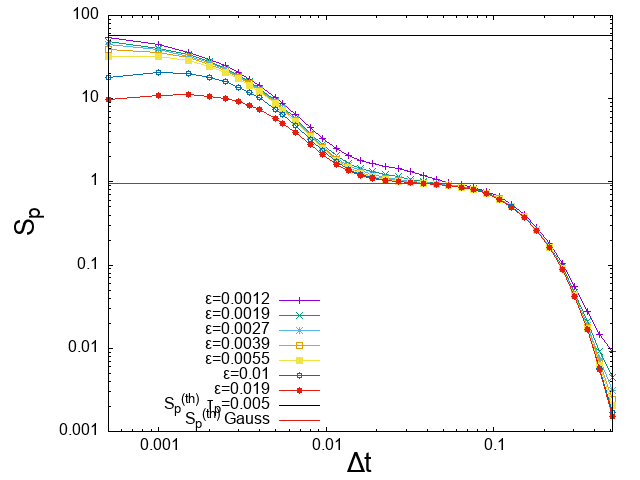}
    \includegraphics[width=0.48\textwidth]{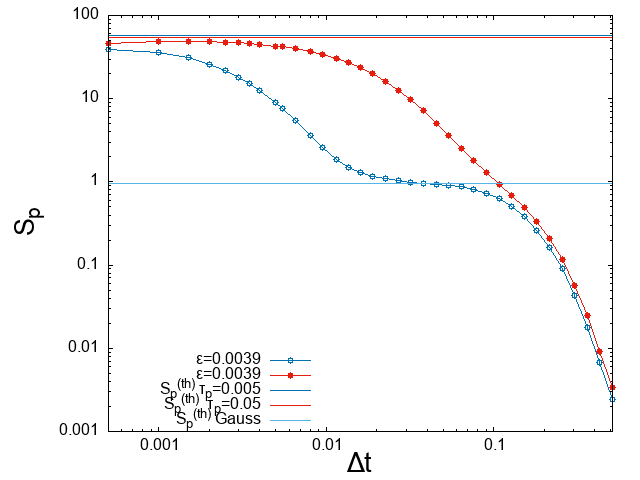}
    \includegraphics[width=0.48\textwidth]{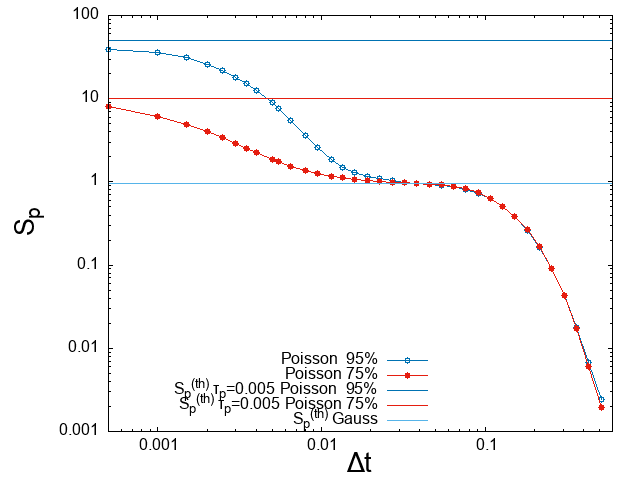}
    \includegraphics[width=0.48\textwidth]{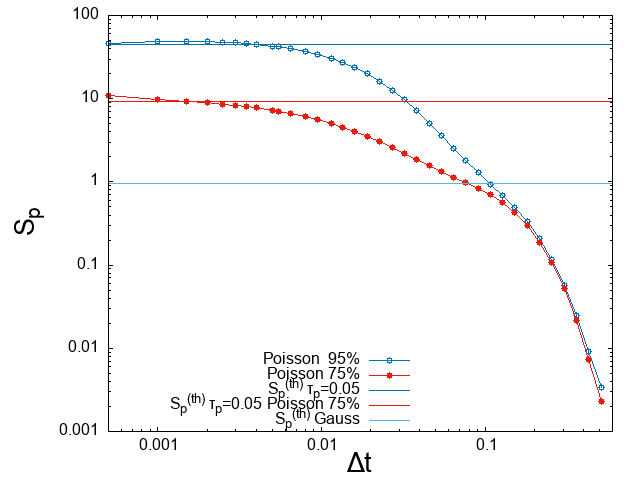}
  \caption{Empirical entropy production rate $\cS(\eps,\Delta t)$ as a function of $\Delta t$ for a particle in a symmetric periodic potential $V_1(X)$ and pulled by a constant force $f$. Top left panel: $\cS(\eps,\Delta t)$ for several $\eps\in[1.2\cdot10^{-3},1.9\cdot10^{-2}]$. Horizontal lines indicate the theoretical value (black) and the Gaussian prediction (red). Top right panel: $\cS(\eps,\Delta t)$ for two different Poisson jump \textcolor{black}{rates} $\tau_p$ ($\tau_p=0.05$ red, $\tau_p=0.005$ blue) for $\eps=3.9\cdot10^{-3}$. Bottom panels show the \textcolor{black}{convergence} of $\cS(\eps,\Delta t)$ for $\eps=3.9\cdot10^{-3}$  towards the theoretical \textcolor{black}{values} (horizontal lines) for different level of Poisson noise ($75\%\,,95\%$) for $\tau_p=0.005$ (left) and $\tau_p=0.05$ (right). The parameters used for numerical simulations are $L=1,V_0=1,d=-0.2,\tau_p=1/\lambda,\sigma^2=0.2,T=(1-p) \sigma^2/2,\Gamma=p\sigma^2/\lambda$, $f=0.5$ with $p=0.75$ or $p=0.95$. \textcolor{black}{The initial conditions are sampled from the stationary distribution.} 
  }
  \label{fig:SpPeriodic_Simple}
\end{figure}
This expectation is confirmed by numerical simulation as can be seen in the top panels of Fig.~\ref{fig:SpPeriodic_Simple}. The left panel shows $\cS(\epsilon,\Delta t)$ as a function of $\Delta t$ for various $\epsilon$. It can be noted that almost every $\epsilon$ \textcolor{black}{displays} a pronounced plateau in the correspondence of the Gaussian prediction. The differences between the Gaussian and non-Gaussian cases only arise for $\Delta t < \tau_p $. Note also that, as explained in the previous section, the convergence towards the true entropy production $\cS$ \textcolor{black}{may not be} very precise due to the large number of samples required. The different regimes of $\cS(\eps,\Delta t)$ are shown on the right panel showing $\cS(\eps,\Delta t)$ for $\tau_p=\{0.005,0.05\}$. Bottom panels of Fig. ~\ref{fig:SpPeriodic_Simple} show $\cS(\eps,\Delta t)$ with two different \textcolor{black}{percentages} of Poisson noise \footnote{\textcolor{black}{The percentage of Poisson noise $p$ is defined as the ratio between the variance of the Poisson noise $\lambda \Gamma$ and the total variance of the noise $\sigma^2=(2T+\lambda \Gamma)$, that is $p=\frac{\lambda \Gamma}{\sigma^2}.$}}. These figures confirmed once again that for large enough $\Delta t$ the coarse-grained entropy production is not sensitive to different noises and  differences only arise when $\Delta t\sim \tau_p$.
\begin{figure}[ht!]
\centering
    \includegraphics[width=0.48\textwidth]{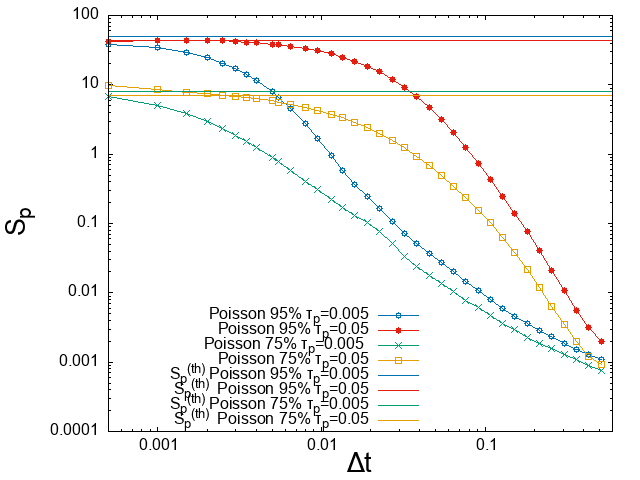}
  \caption{Empirical entropy production rate $\cS(\eps,\Delta t)$ as a function of $\Delta t$ for a particle in an asymmetric periodic potential $V_2(X)$ for $\eps=3.9\cdot 10^{-3}$. Blue (green) and red (yellow) curves show $\cS(\eps,\Delta t)$ for $\tau_p=0.005\,,0.05$ respectively at two different Poisson level $75\%$ (green and yellow) and $95\%$ (blue and red). The parameters used for numerical simulations are $L=1,V_0=1,d=-0.2,\tau_p=1/\lambda,\sigma^2=0.2,T=(1-p) \sigma^2/2,\Gamma=p\sigma^2/\lambda$, $f=0$ with $p=0.75$ or $p=0.95$. \textcolor{black}{The initial conditions are sampled from the stationary distribution.} 
  }
  \label{fig:SpPeriodic_Ratchet}
\end{figure}\\
Most of the \textcolor{black}{considerations} made for the symmetric potential also hold for the ratchet potential. Thus, we focus on the case $f=0$ where a physical current induced by the compound Poisson noise $\zeta$ is present. Here we just want to underline that since the corresponding Gaussian case \textcolor{black}{is} an equilibrium system, for large $\Delta t$ the empirical entropy production $\cS(\eps,\tau)$ drops to $0$ (see Fig.~\ref{fig:SpPeriodic_Ratchet}) although the system \textcolor{black}{sustains} a steady current, as can be seen from Fig.~\ref{fig:Current_Ratchet}. 
\section{The "Poissonian" Gyrator}
\label{sec:bg}
\begin{figure}[ht!]
\centering
    \includegraphics[width=0.48\textwidth]{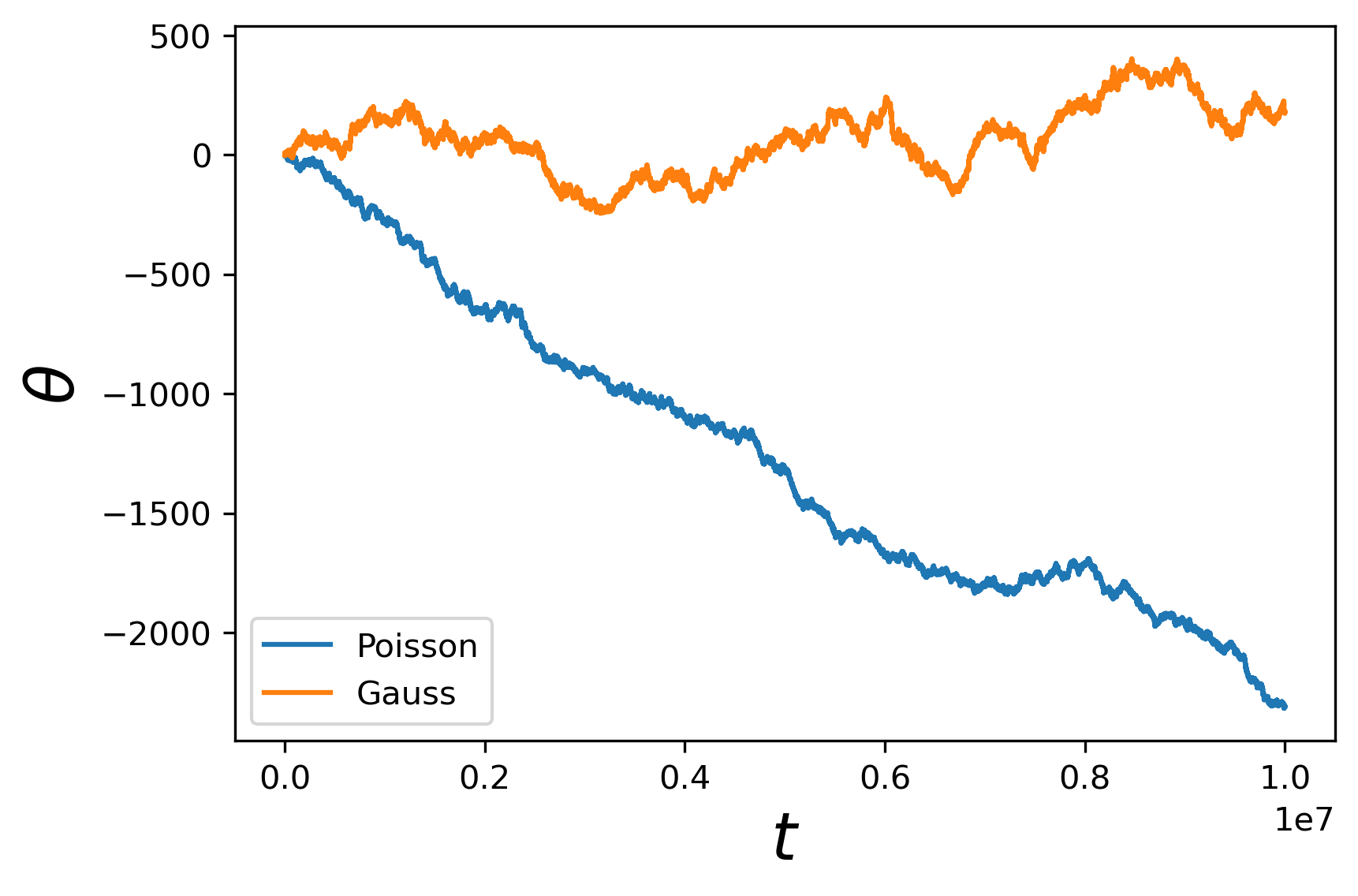}
    \includegraphics[width=0.48\textwidth]{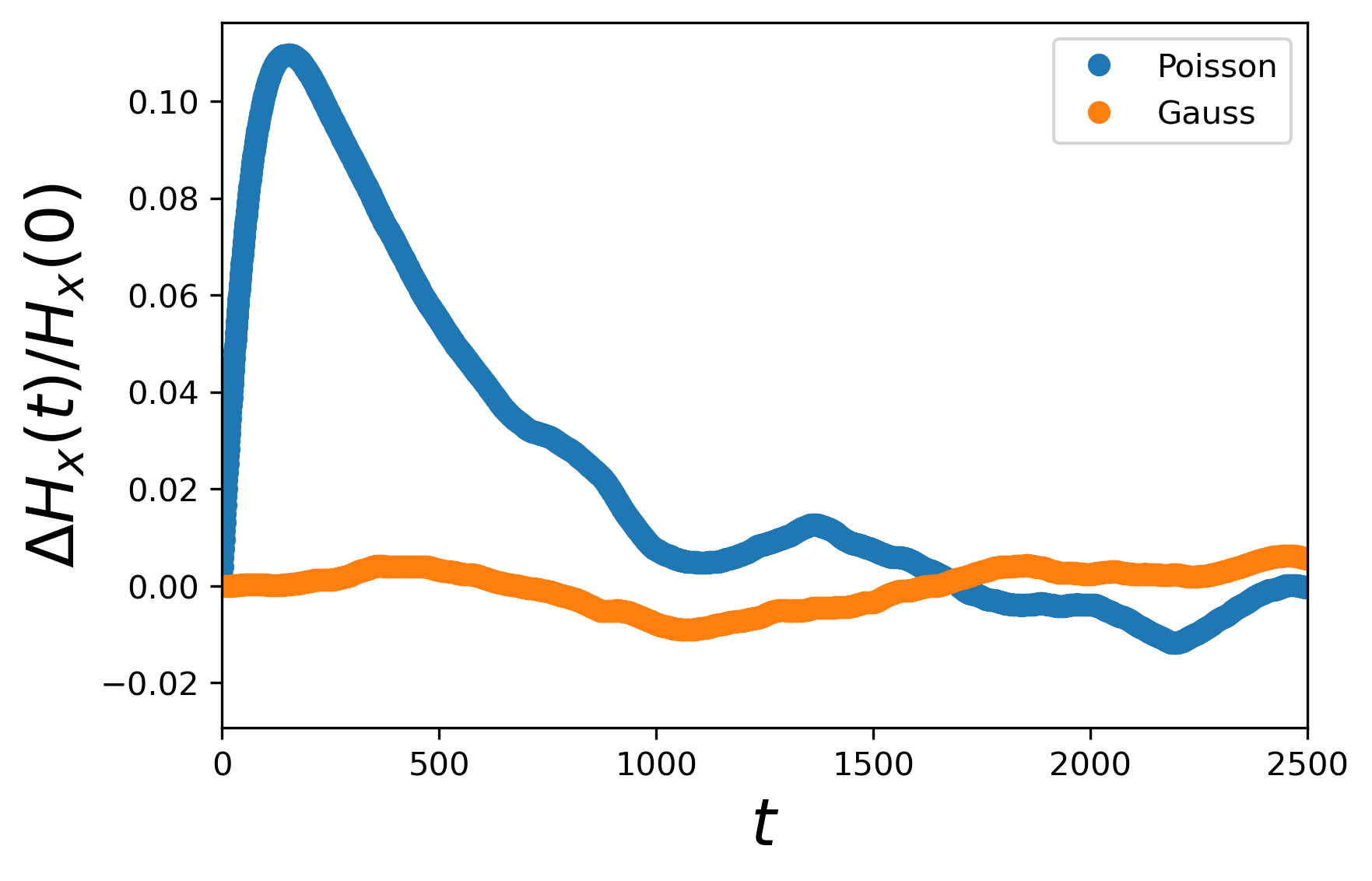}
  \caption{Temporal asymmetries of the Poisson-Brownian Gyrator. Left panel: angle $\theta$ Vs $t$ for the process driven by the compound Poisson noise (blue) and the Gaussian noise orange. The average slope \textcolor{black}{represents} the stationary current $j_s$. Right panel: Degree of irreversibility $\Delta H(t)$ normalized to the value $H(0)$. The degree of irreversibility for the process driven by compound Poisson noise $\zeta$ is shown in blue while orange curve represents the Gaussian case. The parameters used for numerical simulations are $\tau_p=64$, $\lambda=1/\tau_p$, $A=\left(\begin{array}{cc}
  0.00359681 & -0.000899202 \\
-0.000899202 & 0.00226257 \\
    \end{array}\right)$, $\Gamma=\left(\begin{array}{cc}
    0.653682 & 0 \\
    0 & 0 \\
    \end{array}\right)$ and $\Sigma=\left(\begin{array}{cc}
    0 & 0 \\
    0 & 0.01021378 \\
    \end{array}\right)$. \textcolor{black}{The initial conditions are sampled from the stationary distribution.} }
  \label{fig:current_Pomeau_pbg}
\end{figure}

\begin{figure}[ht!]
\centering
    \includegraphics[width=0.48\textwidth]{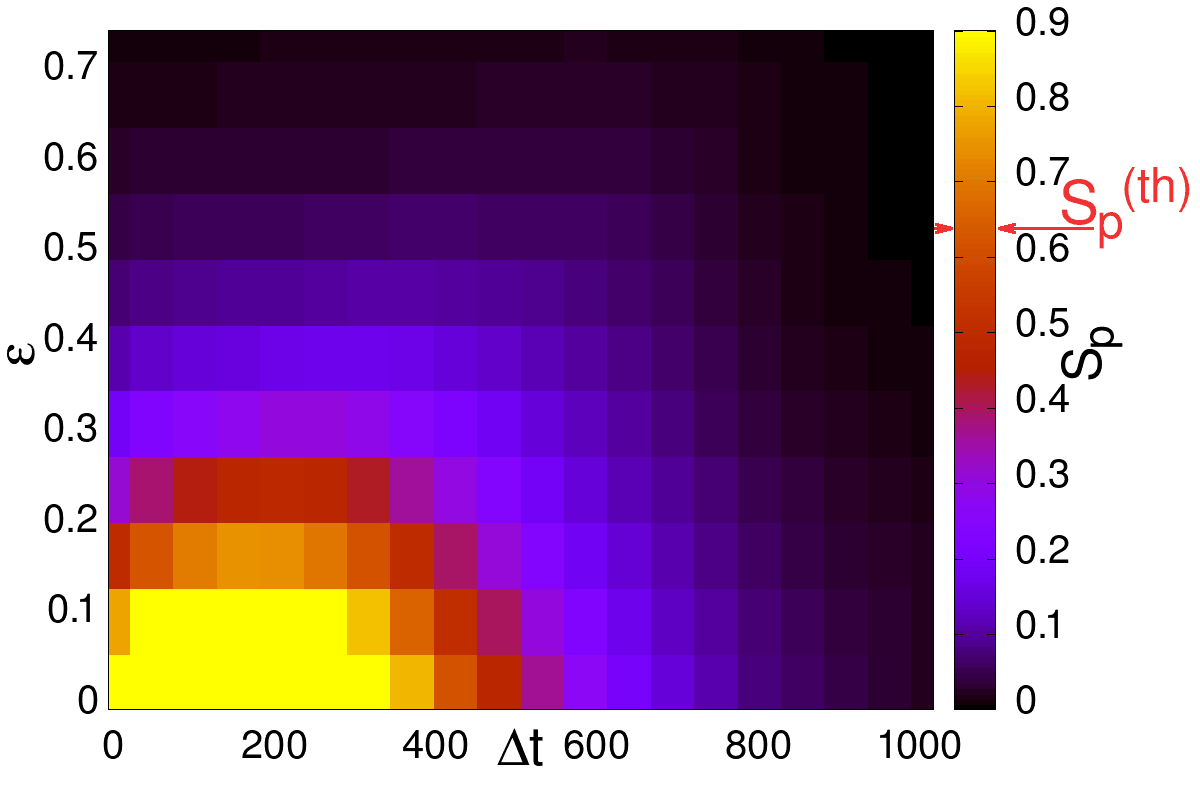}
    \includegraphics[width=0.48\textwidth]{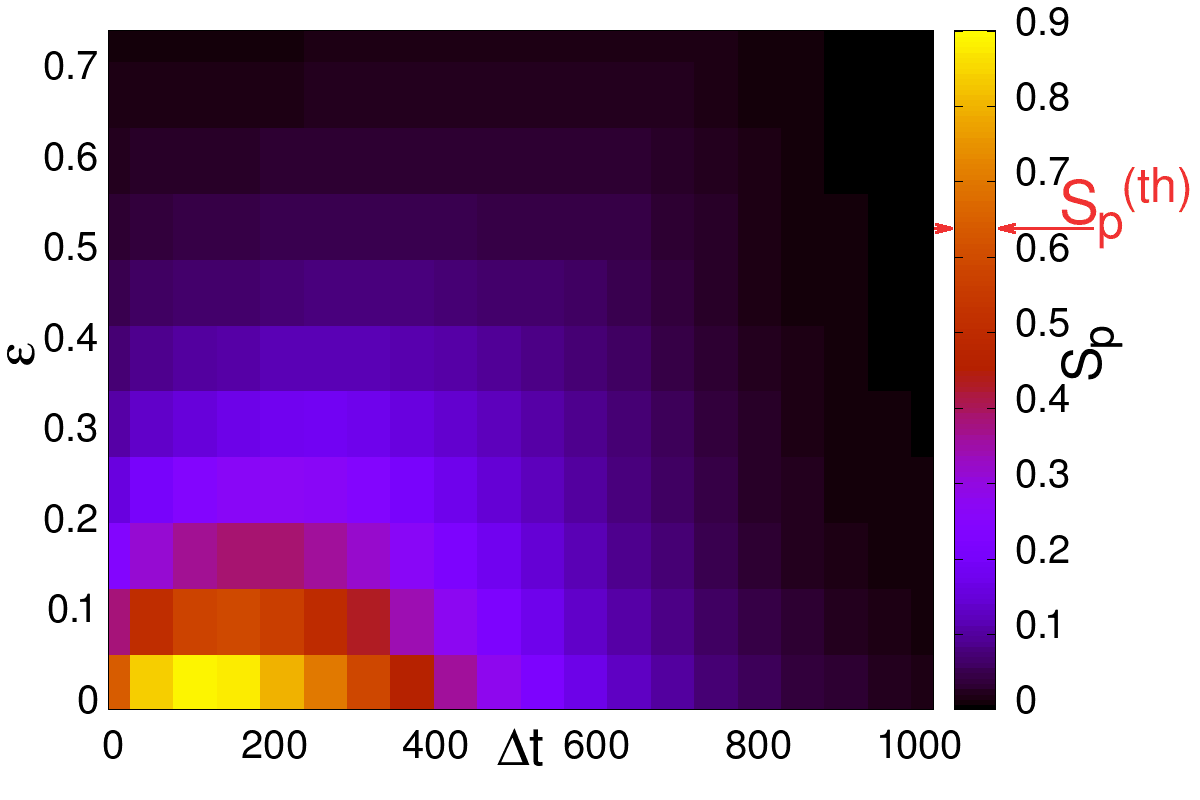}
  \caption{Heat map of empirical entropy production rate $\cS(\eps,\Delta t)$ as a function of $\Delta t$ and $\eps$. Left and right panels show two estimates $\cS(\eps,\Delta t)$ obtained from trajectories of different duration $\cT$ ($\cT=2^{25}$ left, $\cT=2^{28}$ right). The parameters used for numerical simulations are $\tau_p=64$, $\lambda=1/\tau_p$, $A=\left(\begin{array}{cc}
  0.005058  & 0.0012955 \\
-0.00276034 & 0.000801378 \\
    \end{array}\right)$, $\Gamma=\left(\begin{array}{cc}
    0.825827 & 0 \\
    0 & 0 \\
    \end{array}\right)$ and $\Sigma=\left(\begin{array}{cc}
     0.000107637 & 0 \\
    0 & 0.000222583 \\
    \end{array}\right)$. \textcolor{black}{The initial conditions are sampled from the stationary distribution.} }
  \label{fig:SpGirator_Convergence}
\end{figure}

\begin{figure}[ht!]
\centering
    \includegraphics[width=0.48\textwidth]{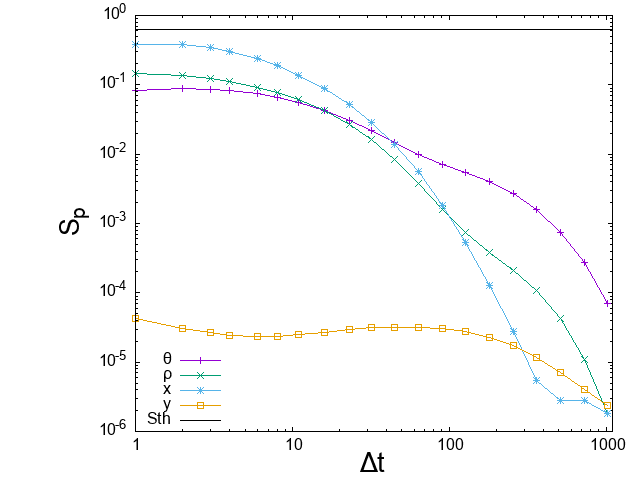}
  \caption{Empirical entropy production rate $\cS(\eps,\Delta t)$ for $\eps=1.2\cdot10^{-2}$ as a function of $\Delta t$ computed from one dimensional signals ($\theta$ purple, $\rho$ green, $x$ light blue, $y$ yellow). The horizontal black line represents the theoretical value. 
  The parameters used for numerical simulations are $\tau_p=64$, $\lambda=1/\tau_p$, $A=\left(\begin{array}{cc}
  0.005058  & 0.0012955 \\
-0.00276034 & 0.000801378 \\
    \end{array}\right)$, $\Gamma=\left(\begin{array}{cc}
    0.825827 & 0 \\
    0 & 0 \\
    \end{array}\right)$ and $\Sigma=\left(\begin{array}{cc}
     0.000107637 & 0 \\
    0 & 0.000222583 \\
    \end{array}\right)$. \textcolor{black}{The initial conditions are sampled from the stationary distribution.} }
  \label{fig:1D_Entropy_pbg}
\end{figure}

In the previous sections we have discussed how the Poissonian noise $\zeta$ affects the equilibrium properties of one dimensional systems. In particular, we have stressed that some equilibrium conditions (such as the absence of physical currents or symmetric correlations) are no longer sufficient to infer thermodynamic properties of the system if the Gaussian fluctuations assumption ceases to hold. However, one-dimensional examples have the drawback that their Gaussian counterpart is necessarily a trivial equilibrium or non-equilibrium system (a forcing is required to sustain non-equilibrium steady states). On the other hand, multidimensional systems can be thrown out of equilibrium without forcing even in the Gaussian case, whenever the system is in contact with multiple thermal baths at different temperatures. Thus, the aim of this section is to present an example of such a system, namely the case of the so-called Brownian gyrator~\textcolor{black}{\cite{filliger2007brownian,ciliberto2013heat,ciliberto2013statistical,argun2017experimental}}. It consists in a two dimensional linear system
\begin{align}
    &\dot{X}=-AX+\xi\,,\\
    &A=\left(\begin{array}{cc}
    \alpha & -\eta \\
    -\mu & \gamma \\
    \end{array}\right)\,,\nonumber\\
    &X=\left(\begin{array}{c}
    x \\
    y \\
    \end{array}\right)
\end{align}
with $\ave{\xi(t)}=0$, $\ave{\xi(t)\xi(t')}=\Sigma\delta(t-t')$ and being $\Sigma=\left(\begin{array}{cc}
    2T_1 & 0 \\
    0 & 2T_2 \\
    \end{array}\right)$. 
In the stationary states, the correlation function takes the form 
\begin{equation}
    \cC(t)=e^{-At}\cC
    \label{eq:correlation_bg}
\end{equation}
where $\cC=\cC(0)$ represents the covariance of $X$. The equilibrium conditions are expressed through the Onsager relations \textcolor{black}{(see \cite{G90,crisanti2012nonequilibrium,lucente2022inference} or~\ref{app:Linear-System})}
\begin{equation}
    A\cC=\cC A^T
\label{eq:onsager}
\end{equation}
or equivalently from the conditions $\cC(t)=\cC^T(t)$. When Eq. (\ref{eq:onsager}) is not satisfied, a systematic torque induces a rotational current ($\langle\dot{\theta}\rangle\neq 0$ with $\theta=\arctan\lx(\frac{y}{x}\rx)$) in the system~\cite{crisanti2012nonequilibrium,baldassarri2020engineered}. In these cases, the entropy production \textcolor{black}{rate} $\cS$ is 
\begin{equation}
    \cS=\text{Tr}\lx[2 \cC A^T \Sigma^{-1}A - A \rx]=\frac{\lx(T_2\eta-T_1\mu\rx)^2}{T_1T_2(\alpha+\gamma)}
    \label{eq:epr_bg}
\end{equation}
which vanishes when $T_2\eta=T_1\mu$. Furthermore, the rotational current $\langle\dot{\theta}\rangle$ is proportional to  $\lx(T_2\eta-T_1\mu\rx)$ and is equal to zero in equilibrium conditions. \textcolor{black}{It should also be noted that when the system is coupled to two different heat baths ($T_1\neq T_2$) the interaction terms $\mu$, $\eta$ must be non-reciprocal ($\mu\neq\eta$) in order to maintain equilibrium~\cite{loos2020irreversibility}.} As already anticipated in Sec.~\ref{sec:ts} and analogously to the one-dimensional examples, the presence of an additional Poissonian noise $\zeta$ changes the above picture. Thus, consider the equation
\begin{equation}
    \dot{X}=-AX+\xi\,+\zeta
\end{equation}
where, as usual, we denote the covariance of jumps $U$ as $\Gamma$. The correlations of the process in the stationary state read
\begin{equation}
\hat{\cC}(t)=e^{-At}\hat{\cC}
\label{eq:correlation_pbg}
\end{equation}
with $\hat{\cC}=\hat{\cC}(0)$. Note that Eqs. (\ref{eq:correlation_pbg}) are equivalent to Eqs. (\ref{eq:correlation_bg}) and so $\hat{\cC}$ has the same structure of $\cC$ of a Gaussian process with noise matrix $D=\Sigma+\lambda\Gamma$ (see \textcolor{black}{\ref{app:Linear-System}} for further details). \textcolor{black}{Since the system is driven by Poisson noise $\zeta$ and it is necessarily out of equilibrium, the previous observation means that Onsager relations \cite{onsager1931} fail to discriminate equilibrium and non-equilibrium systems. For the same reason, also the Harada-Sasa equality that relates the energy dissipation to the violation of Fluctuation-Response relation \cite{harada2005equality} it is not able to distinguish between equilibrium and non-equilibrium dynamics}~\footnote{\textcolor{black}{The Harada-Sasa equality is \cite{harada2005equality} $$J=\gamma\int{{\rm d}\omega\,\lx[\tilde{C}(\omega)-2T \tilde{R}'(\omega)\rx]}$$ where $J$ is the rate of energy dissipation, $\tilde{C}(\omega)$ and $\tilde{R}(\omega)$ are the Fourier transforms of the correlation function $C(t)$ and the response $R(t)$, and the prime denotes the real part. }}. Interestingly, in "equilibrium-like" conditions, namely $A\hat{\cC}=\hat{\cC} A^T$, the system displays a non-vanishing rotational current $\langle \dot{\theta}\rangle\neq 0$, as one can verify in the left panel of Fig.~\ref{fig:current_Pomeau_pbg} which shows the cumulated $\theta$ angle, i.e. $\int_0^t ds \dot\theta(s)$, as a function of time. Thus, the Poisson-Brownian gyrator behaves as a linear Brownian ratchet embedded in two dimensions. It should be noted that it is possible to find a stall condition ($\langle \dot{\theta} \rangle=0$) imposing different temperatures of the Gaussian baths ($T_1\neq T_2$) so that a vanishing current does not imply anymore equilibrium. The non-equilibrium nature of the system can always be inferred from higher-order \textcolor{black}{correlation functions} as for instance from $H_x(t)=\ave{x(t)x^3(0)}$, as can be seen from right panel of Fig.~\ref{fig:current_Pomeau_pbg} which shows $\Delta H_x$ for both the Gaussian and Poissonian cases.  
\\
Although in this case it is not possible to use energetic arguments for estimating the entropy produced between two subsequent jumps, thanks to the linearity of the system it is still possible to perform analytical computations, as detailed in \textcolor{black}{\ref{app:Linear-System}}. It turns out that the entropy production \textcolor{black}{rate} $\cS$ is given by
\begin{equation}
    \cS=\text{Tr}\lx[2 \hat{\cC} A^T \Sigma^{-1}A - A \rx]
    \label{eq:epr_pbg}
\end{equation}
which is exactly Eq. (\ref{eq:epr_bg}) with $\cC$ replaced by $\hat{\cC}$. It is important to stress that the "temperatures" appearing in the denominators are those of the Gaussian baths only. If one splits the correlation matrix in two parts $\hat{\cC}=\cC+\cC'$ related to the Gaussian and Poissonian covariance matrices respectively, the entropy production \textcolor{black}{rate} $\cS$ can be written as
\begin{equation}
    \cS=\text{Tr}\lx[2 \cC A^T \Sigma^{-1}A - A \rx]+2\lambda \text{Tr}\lx[\cC' A^T \Sigma^{-1}A\rx]
    \label{eq:epr_pbg_split}
\end{equation}
where the first term is the usual entropy production of the Gaussian system while the second term (always positive) is the contribution of the Poisson jump noise. When Onsager relations ($A\hat{\cC}=\hat{\cC} A^T$) are satisfied, the entropy production \textcolor{black}{rate} attains its minimum, that is 
\begin{equation}
\cS=\lambda \text{Tr}\lx[\Gamma \Sigma^{-1}A\rx] \,.
\end{equation}
Interestingly, the minimum is not unique and $\cS=\lambda \text{Tr}\lx[\Gamma \Sigma^{-1}A\rx]$ whenever $A\Sigma=\Sigma A^T$, as detailed in \textcolor{black}{\ref{app:Linear-System}}. 
Having \textcolor{black}{an} analytical formula for the entropy production \textcolor{black}{rate}, it is natural to wonder how its empirical estimates behave in this case. Obviously, in the analysis of the results it is necessary to keep in mind the limitations that have arisen in the one-dimensional cases. In particular, in Sec.~\ref{sec:epr_linear_1d}, we have shown that it often takes a gigantic amount of data to obtain reasonable estimates of $\cS$. In multidimensional systems this problem is accentuated and therefore we expect it to impose severe limitations on the achievable resolutions. In the following we consider situations where the Poissonian noise acts only on one component ($x$). The panels of Fig.~\ref{fig:SpGirator_Convergence} show the empirical entropy production $\cS(\eps,\Delta t)$ as a function of both $\epsilon$ and $\Delta t$ for two different trajectories duration ($\cT=2^{25}$ left, $\cT=2^{28}$ right). 
From both figures it can be seen that for time scales small enough ($\Delta t \sim 100$) and spatial scales small but large enough to have suitably large statistics ($\eps\sim 0.1$), the estimates of entropy production \textcolor{black}{rate} are in good agreement with the theoretical predictions. Furthermore, comparing left and right panels it should be noted that as the samples size increases the convergence of the estimates $\cS(\eps,\Delta t)$ towards the analytical prediction also improve. \\
To conclude, we discuss the behavior of empirical estimates $\cS(\eps,\Delta t)$ when only partial information about the system is available. For example, consider the case in which the time series of a single scalar variable has been observed. 
There is growing interest around this topic, mainly due to the large number of thermodynamic bounds on entropy production recently proposed relying both on thermodynamic uncertainty relations (TURs) (see~\cite{barato2015thermodynamic,seifert2018stochastic,plati2023thermodynamic,di2023variance} and reference therein) or on other quantities such as dynamical activity (also known as "frenesy")~\cite{maes17,terlizzi18}.
Generally, TURs provide a lower bound of entropy production as a ratio between average and fluctuations of a steady current. Thus, \textcolor{black}{TURs} strongly differ from coarse-grained entropy production  $\cS(\eps,\Delta t)$ since the latter can provide non-trivial estimates, even if the system does not sustain any steady physical current, as shown in Sec.~\ref{sec:ep}. In the Poisson-Brownian gyrator, there are some natural one-dimensional variables among all possible one dimensional signals: the radius $\rho$, the angle $\theta$ and the two components $x$ and $y$. Note that only the angle $\theta$ can be used in a TUR to provide a lower bound on entropy production since it is the only variable which can display a current. For the other variables, the best we can do is to study the empirical estimate $\cS(\eps,\Delta t)$. Fig.~\ref{fig:1D_Entropy_pbg} shows these \textcolor{black}{estimates} as a function of $\Delta t$ for a given value of the spatial resolution $\epsilon$. 
The first thing to note is that the estimates obtained from the $y$ signal are practically zero. This is due to the fact that Poisson noise $\zeta$ acts indirectly on $y$ through the coupling with $x$. The best estimate $\cS(\eps, \Delta t)$ is obtained considering the signal $x$ since it feels directly the jump noise $\zeta$. The estimates provided by the signals $\rho$ and $\theta$ are in between those provided by $x$ and $y$ separately as $\rho$ and $\theta$ are non-linear combination of these two signals. This analysis shows that in general the bounds of entropy productions on partially observed systems strongly depend on the variables considered and therefore usually can not be considered good proxies of the true value. This claim is supported by Tab.~\ref{tab:1d_estimates_asymptotics} showing the ratio between empirical estimates and theoretical value. As can be noted, from $x$ and $\rho$ one gets estimates of the same order of magnitude of the theoretical value ($60\%$ and $20\%$ respectively). On the other hand, the estimate provided by the signal $y$ is $10^4$ times smaller than the correct value. Interestingly, the signal $\theta$ provides an estimate which is one order of magnitude smaller than the theoretical value if one takes as empirical estimate the coarse-grained entropy production $\cS(\eps,\Delta t)$. However, if one employs the TUR\footnote{Formally the TUR is defined as $\text{TUR}=\lim_{t\to\infty}\frac{m^2(t)}{t\sigma^2(t)}$ where $m(t)=\ave{\theta_t-\theta_0}$ and $\sigma^2(t)=\ave{(\theta_t-\theta_0)^2}-m^2(t)$~\textcolor{black}{\cite{barato2015thermodynamic,seifert2018stochastic}}.} to provide a lower bound \textcolor{black}{he gets a value which is two order of magnitude smaller than the theoretical one.} \textcolor{black}{Therefore, in the case of discontinuous processes, we argue that TURs, although technically valid, usually severely underestimate the entropy production.} This means that estimates of entropy production make sense when one has a clear understanding of the model suitable for describing a phenomenon, otherwise the results strongly depend on the chosen model.
 
\begin{table}[h]
\centering
\begin{tabular}{|c|c|}
\hline
 Signal & $\cS(\eps_{min},\Delta t_{min})/\cS$  \\
\hline
$x$  & 0.58 \\
$y$  & $0.64\cdot 10^{-4}$  \\
$\rho$ & 0.23\\
$\theta$ & 0.12\\
\hline
TUR & $9.72\cdot 10^{-3}$ \\
\hline
\end{tabular}
\label{tab:1d_estimates_asymptotics}
\caption{Table with the ratio of the empirical estimate of entropy production $\cS(\eps,\Delta t)$ and the theoretical value for different one dimensional signals. The last row represents the estimate obtained with TUR.}
\end{table}
\section{Conclusions}
In this work we have thoroughly discussed the role of non-Gaussian white noise, as a driving force, on the non-equilibrium properties of a system. In particular, it has been shown that Langevin equations driven by symmetric Poissonian noises do not satisfy detailed balance and are therefore inherently out-of-equilibrium. However, non-equilibrium manifests itself quite differently than in systems driven by Gaussian noise and a plethora of unusual behaviors can be observed. For example, the absence of currents or the symmetries of correlation functions do not guarantee anymore that the system is in thermodynamic equilibrium. We have shown that, from an experimental point of view, it is rather easy to evidence the breaking of time reversal symmetry by considering higher-order correlation functions. Nonetheless, these \textcolor{black}{methods} cannot provide an intrinsic measure of the degree of irreversibility of a given system. Formally, the difficulties are overcome by considering an information-theoretic quantity, namely the entropy production \textcolor{black}{rate} $\cS$. We show that it is possible to provide explicit \textcolor{black}{formulas} for the path probabilities of stochastic processes driven by both Gaussian and Poissonian noise but unfortunately analytical expressions for $\cS$ can be obtained in special cases only. These cases play an important role because they underline that the "temperatures" of the thermal baths in which the system dissipates are exclusively the Gaussian ones. Thus, the entropy production diverges when the system is not coupled to Gaussian thermal baths justifying the expression "athermal" baths already used for Poissonian noise~\cite{kanazawa2015minimal,kanazawa2017statistical}. \\
In addition to the analytical results, an empirical estimate of the entropy production has  been introduced, i.e. the $\epsilon-\Delta t$ entropy production $\cS(\eps,\Delta t)$ depending on the spatial and temporal resolution scales. Numerical simulations provide clear evidence of the goodness of this approach, although it often requires an immense amount of data to converge to the analytical prediction. It therefore can not always be applied to the analysis of experimental signals, especially in high-dimensional systems, since the finite number of samples could lead to very inaccurate results. It is also worth noting that for low temporal resolution a system may appear indistinguishable from one driven by Gaussian noise. The "Poissonian" gyrator is an explanatory example of the behavior of this type of systems, revealing itself as an excellent test bed for discussing the estimates of entropy production from incomplete information. We have shown that the estimates strongly depend on the chosen observables, varying even by four orders of magnitude with respect to the theoretical value. It can therefore be concluded that without a deep comprehension of the system under consideration, it is very difficult to provide accurate estimates of its thermodynamics properties.

\label{sec:concl}
\section*{Acknowledgments}
We thank M. Baldovin for helpful discussions and for his accurate reading of the manuscript. A.P. and A.V. acknowledge the financial support from the MIUR PRIN 2017 (project "CO-NEST" no. 201798CZLJ). M.V. and D.L. were supported by ERC-2017-AdG (project "RG.BIO" no. 785932). M.V. was supported by MIUR FARE 2020 (project "INFO.BIO" no. R18JNYYMEY).

\appendix
\section{The Setting}\label{app:Setting}
\textcolor{black}{In this section we are interested in discussing more in detail the non-equilibrium properties of stochastic processes introduced in Sec.~\ref{sec:ts}, i.e. stochastic processes in which random jumps are added to a typical Wiener process.} 
The dynamic of such processes can be formally described by the following stochastic differential equation
\begin{equation}
    \dot{X} = F(X) + \xi(t) + \zeta(t) \qquad X=\{x_i\}_{i=1,N}
\end{equation}
where $\xi(t)$ is a standard Gaussian noise with $\ave{\xi_i(t)}=0$ and $\ave{\xi_i(t)\xi_j(t')}=\Sigma_{ij}\delta(t-t')$ while $\zeta(t)=\sum_j U_j \delta(t-t_j)$ where both the intervals between to subsequent jumps $t=t_{j}-t_{j-1}$ and the amplitude of jumps $U$ are i.i.d. drawn \textcolor{black}{from} the probability distributions $\cQ_\lambda(t)$ and $\cP(U)=\cG_\Gamma(U)$ defined as
\begin{eqnarray}
    &\cQ_\lambda(t) = \lambda e^{-\lambda t}\\
    & \cG_\Gamma(U) = \frac{e^{-\frac{1}{2}U^T\Gamma^{-1}U}}{\sqrt{|2\pi\Gamma|}}\,.
\end{eqnarray}
Once the jumps in the time interval $[0,t)$ have been extracted (we assume that there have been $n$ jumps at times $0<t_1<t_2<\dots<t_n<t$ of intensity $U_1,U_2,\dots,U_n$ and we denote this set of values with $\cK_n\equiv \{t_k,U_k\}_{k=1,n}$), since between two jumps we have a Wiener process, we can write down the transition probability $\cW^{(n)}_t(X|Y,\cK_n)$ from $X(0)=Y$ to $X(t)=X$ of the whole process simply by suitably concatenating the "free" propagators $\cW^{(0)}_t(X|Y)$ of the Wiener process , those we have in absence of jumps ($\cK_0=\emptyset$), i.e. 
\begin{equation}
\cW^{(n)}_t(X|Y,\cK_n) = \int \prod_{k=1}^n \lx[ dX_k \, \cW^{(0)}_{t_k - t_{k-1}}( X_k - U_k|X_{k-1}) \rx] \cW^{(0)}_{t-t_n}(X|X_n)
\end{equation}
where $X_0=Y$ and $t_0=0$.
The expression above is the starting point from which we can deduce the non-equilibrium properties of such processes once we figure out how to average over the jump distribution $\cP(\cK_n)$.
\section{Averaging Over the Jump Distribution}\label{app:Notation}
\textcolor{black}{The aim of this section is to explain how to average generic functions over the jump distribution. On one hand these calculations serve to clarify how to correctly define the path probability \textcolor{black}{that} appeared in Sec.~\ref{sec:ep}, while on the other they are a necessary step for the computation of the entropy production carried out in the following appendices.}
First of all, since the distribution of the time intervals $t_k - t_{k-1}$ between two jumps is exponential, the probability of having $n$ jumps in a time $t$ is Poissonian, i.e. $\cP_\lambda(n|t) = e^{-\lambda t}(\lambda t)^n/n!$, while, given $n$ and $t$, the distribution $\cP(t_1,\dots,t_n|n,t)$ of the times $\{t_k\}_{k=1,n}$ in the interval $[0,t)$ is 
\begin{equation}
\cP(t_1,\dots,t_n|n,t) = 
\begin{cases}
n!/t^n & 0\leq t_1<t_2<\dots<t_n<t \\
0 & \text{otherwise}
\end{cases}
\end{equation}
Putting it all together we get
\begin{equation}
\cP(\cK_n|t) = \cP_\lambda(n|t) \cP(t_1,\dots,t_n|n,t) \prod_{k=1}^n \cG_\Gamma(U_k) 
\end{equation}
with
\begin{equation}
\sum_{n=0}^\infty \int d\cK_n \cP(\cK_n|t) = 1 \quad \lx( dK_n = \prod_{k=1,n} d t_k dU_k\, \forall n>0,\quad\cP(\cK_0|t)=e^{-\lambda t}\rx)
\end{equation}
Now, we are able to compute the average $\overline{\cF_t}=\sum_{n=0}^\infty \int d\cK_n \cP(\cK_n|t) \cF_t(\cK_n)$  of a generic function of the type 
\begin{equation}
    \cF_t(\cK_n) = \sum_{k=1}^n \cF(t-t_k,U_k) \quad \forall n>0
\end{equation}
In fact, given $t$, if we assume that in the absence of jumps such function is $\cF_t(\cK_0)=\cF^{(0)}_t$, we have
\begin{equation}
    \overline{\cF_t} - e^{-\lambda t}\cF^{(0)}_t = e^{-\lambda t} \sum_{n=1}^\infty \lambda^n \sum_{k=1}^n \int_0^t d t_1 \int_{t_1}^t d t_2 \dots \int_{t_{k-1}}^t d t_k\, f(t-t_k) \frac{(t-t_k)^{n-k}}{(n-k)!}
    = e^{-\lambda t} \cI^f(t)
\end{equation}
where $f(t) = \int dU \cG_{\Gamma}(U) \cF(t,U)$, $\cI^f(t) = \sum_{n=1}^\infty \cI^f_n(t)$, $\cI^f_n(t) = \lambda^n \sum_{k=1}^n \cI^f_{n,k}(t)$ and~\footnote{Note that the factor $t^n/n!$ is canceled by its inverse present in the probability of having $n$ jumps while factor $(t-t_k)^n/(n-k)!$ is obtained by integrating \textcolor{black}{over} the values $dt_{k+1}\cdots dt_n$.}
\begin{align}
    \cI^f_{n,k}(t) &= \int_0^t d t_1 \int_{t_1}^t d t_2 \, \dots \, \int_{t_{k-1}}^t d t_k \, f(t-t_k) \frac{(t-t_k)^{n-k}}{(n-k)!} \nonumber\\ 
    &= \int_0^t d z_1 \int_0^{z_1} d z_2 \, \dots \, \int_0^{z_{k-1}} d z_k \, f(z_k) \, \frac{z_k^{n-k}}{(n-k)!} 
\end{align}
If we differentiate $\cI^f_{n,k}(t)$ with respect to $t$ we get
\begin{equation}
\dot{\cI}^f_{n,k}(t)=\cI_{n-1,k-1}(t) \quad \forall, n,k>1
\end{equation}
and then
\begin{eqnarray}
    \dot{\cI}^f(t) &=& \sum_{n=1}^\infty \dot{\cI}^f_n(t) =  \dot{\cI}^f_1(t) + \sum_{n=2}^\infty \lambda^n \lx(\dot{\cI}^f_{n,1}(t) + \sum_{k=2}^n \cI^f_{n-1,k-1}(t)\rx) = \nonumber\\
    &=& \dot{\cI}^f_1(t) + \sum_{n=2}^\infty \lambda^n \dot{\cI}^f_{n,1}(t) + \lambda \cI^f(t) 
\end{eqnarray}
So, since $\dot{\cI}^f_{n,1}(t) = f(t) t^{n-1}/(n-1)!$, we have to solve the following linear differential equation
\begin{equation}
    \dot{\cI}^f(t) -\lambda \cI^f(t) = \lambda e^{\lambda t} f(t) \qquad \text { with } \cI^f(0) = 0
\end{equation}
in order to get
\begin{equation}
\overline{\cF_t} = e^{-\lambda t}\cF^{(0)}_t + \lambda \int_0^t dt'\, f(t')
\label{avejumpsum}
\end{equation}
We can follow a similar approach in order to compute the average over the jumps for a function like
\begin{equation}
    \cF_t(\cK_n) = \prod_{k=1}^n \cF(t-t_k, U_k) \quad \forall n > 0
\end{equation}
In this case we have
\begin{equation}
 \overline{\cF_t} - e^{-\lambda t}\cF^{(0)}_t = e^{-\lambda t} \sum_{n=1}^\infty \lambda^n \int_0^t d t_1 f(t-t_k) \int_{t_1}^t d t_2 f(t-t_2) \dots \int_{t_{k-1}}^t d t_n\, f(t-t_n) = e^{-\lambda t} \cI^f(t)
\end{equation}
where, again, $f(t) = \int dU \cG_{\Gamma}(U) \cF(t,U)$, $\cI^f(t)=\sum_{n=1}^\infty \cI^f_n(t)$, 
\begin{equation}
\cI^f_n(t) = \lambda^n \int_0^t dx_1 f(x_1) \int_0^{x_1} dx_2 f(x_2) \dots  \int_0^{x_{n-1}} dx_n f(x_n)\end{equation}
and their derivatives respect to $t$ read
\begin{eqnarray}
\dot{\cI}^f_n(t) &=& \lambda f(t) \cI^f_{n-1}(t) \quad \forall n>1 \qquad \lx(\dot{\cI}^f_1(t) = \lambda f(t)\rx)\\
\dot{\cI}^f(t) &=& \lambda f(t) + \lambda f(t) \cI^f(t)
\end{eqnarray}
Once we solve the differential equation above ($\cI^f(0)=0$) we get
\begin{equation}
    \overline{\cF_t} = e^{-\lambda t} \lx(\cF^{(0)}_t + \exp\lx\{\lambda \int_0^t dt' f(t')\rx\} - 1\rx)
\label{avejumppro}
\end{equation}
\subsection{Averaging Over the Stationary Measure}
Given the starting point $X(0)=Y$ and the set of jumps in the time interval $[0,t)$ we define the average $\ave{f(X)|Y,\cK_n}^{(n)}_t$ of a generic function $f(X)$ over the transition probability $\cW^{(n)}_t(X|Y,\cK_n)$ as
\begin{equation}
\ave{f(X)|Y,\cK_n}^{(n)}_t = \int dX\, \cW^{(n)}_t(X|Y,\cK_n) f(X)
\end{equation}
Once we average over the jump distribution also, i.e.
\begin{equation}
\overline{\ave{f(X)|Y}_t} = \sum_{n=0}^\infty \int d\cK_n\, \cP(\cK_n|t) \ave{f(X)|Y,\cK_n}^{(n)}_t
\end{equation}
we can look at the $t\to\infty$ limit to get the average over the stationary measure of the process, i.e.
\begin{equation}
 \overline{\ave{f(X)}} = \int dX \pi(X) f(X) = \lim_{t\to\infty} \overline{\ave{f(X)|Y}}_t
 \end{equation}
where
\begin{equation}
    \pi(X) = \lim_{t\to\infty} \sum_{n=0}^\infty \int d\cK_n\, \cP(\cK_n|t)  \cW^{(n)}_t(X|Y,\cK_n)
\end{equation}
\section{Entropy Production Rate}\label{appendix:EPR}

In Sec.~\ref{sec:ep} the entropy production $\cS$ of a Markov process $X$ has been defined as 
\begin{equation}
    \cS=\lim_{\cT\to\infty}\frac{1}{\cT}\ave{\log\lx(\frac{\cP(\{X_t\}_{0\le t \le \cT})}{\cP(\{ X_{\cT - t}\}_{0 \le t \le \cT})}\rx)}=\lim_{\cT\to\infty} \frac{\ave{\cS_\cT}}{\cT}
\end{equation}
but it has also been mentioned that it can be equivalently defined as 
\begin{equation}
\cS=\lim_{t\to 0} \frac{1}{t}\sum_{X,Y}\pi(X)\cW_t(Y|X)\log\lx(\frac{\cW_t(Y|X)}{\cW_t(X|Y)}\rx)
\label{eq:epr_zero_limit}
\end{equation}
where $\cW_t(Y|X)$ denotes the propagator of the Markovian dynamics.
Indeed, let $h$ be a fixed time step, $t$ the duration of a trajectory and consider the quantities
\begin{eqnarray}
    \cS_t^{(h)} &=& \int dx_0\, \pi(x_0) \prod_{i=1}^N \lx[dx_i\, \cW_h(x_{i-1} \to  x_i) \rx] 
    \lx\{ \log{\frac{\pi(x_0)}{\overleftarrow{\pi}(x_n)}} + \sum_{j=0}^N \log{\frac{\cW_h(x_{j-1} \to  x_j)}{\overleftarrow{\cW_h}(x_j \to  x_{j-1})}}\rx\} \nonumber\\
    &=& \int dx\, \pi(x) \log{\frac{\pi(x)}{\overleftarrow{\pi}(x)}} + \frac{t}{h} \int dxdy\, \pi(x) \cW_h(x\to y) \log{\frac{\cW_h(x \to y)}{\overleftarrow{\cW_h}(y \to  x)}} \nonumber \\
    &=& \cS_0 + t \cS_1(h)
\end{eqnarray}
and 
\begin{eqnarray}    
    \cS_2(t) &=& \int dx dy \cP_t(x,y) \log{\frac{\cP_t(x,y)}{\overleftarrow{\cP_t}(x,y)}} = \cS_0 + t \cS_1(t)
\end{eqnarray}
where
\begin{eqnarray}
    &t = Nh \\
    &\cS_0 = \int dx\, \pi(x) \log{\frac{\pi(x)}{\overleftarrow{\pi}(x)}} \\
    &\cS_1(h) = \frac{1}{h} \int dx \, \pi(x) \int dy \, \cW_h(x\to y) \log{\frac{\cW_h(x\to y)}{\overleftarrow{\cW_h}(y \to x)}}
\end{eqnarray}
with $\overleftarrow{\cdot}$ representing the backward dynamics.\\
Defining $\cS_1 = \lim_{h \to 0} \cS_1(h)$ one has
\begin{eqnarray}
    &\cS_t = \cS_0 + t \cS_1 \\
    &\lim_{t\to\infty} \frac{\cS_t}{t} = \cS_1 = \partial_t \cS_t
\end{eqnarray}
proving that the two definitions of entropy production coincide. However, when the process is discontinuous, the propagator also depends on the exact realizations of the jumps, as can be seen from Eq.~\ref{eq:path_probability}. Thus, at each time step $h$, the system evolves according  the propagator $\cP_h(x \to y)=\cW^{(n)}(x \to y)$ depending on the number of jumps $n$ occurred in a time $h$. Therefore, in order to compute the entropy production $\cS$ employing Eq.~\ref{eq:epr_zero_limit} we should consider the contribution to $\cS$ coming from all trajectories having $n$ jumps and then average over the jump distribution.\\ 
To summarize the previous discussion, for computing the entropy production rate of such processes we have to:
\begin{enumerate}
\item consider the propagator of the reverse path for which we back-jump the system using exactly the same times and intensities, i.e.
\begin{equation}
\overleftarrow{\cW}^{(n)}_t(Y|X,\cK_n) = \int \prod_{k=1}^n \lx[ dX_k \, \cW^{(0)}_{t_k - t_{k-1}}( X_{k-1}|X_k-U_k ) \rx] \cW^{(0)}_{t-t_n}(X_n|X)
\end{equation}
\item compute the entropy production rate once $t$ and the jumps $\cK_n$ are given
\begin{equation}
\cS^{(n)}_t(Y,\cK_n) = \frac{1}{t} \int dX \cW^{(n)}_t(X|Y,\cK_n) \log\frac{\cW^{(n)}_t(X|Y,\cK_n)}{\overleftarrow{\cW}^{(n)}_t(Y|X,\cK_n)},
\end{equation}
\item average over the jump distribution
\begin{equation}
\cS_t(Y) = \sum_{n=0}^\infty \int d\cK_n\, \cP(\cK_n) \cS^{(n)}_t(Y,\cK_n),
\end{equation}
\item remove the dependence on the starting point $Y$ by averaging over the stationary measure $\pi(Y)$
\begin{equation}
\cS_t = \int dY\, \pi(Y) \cS_t(Y)
\end{equation}
\item take the $t\to 0$ limit
\begin{equation}
\cS = \lim_{t\to0} \cS_t
\end{equation}
\end{enumerate}
\section{Linear Systems}\label{app:Linear-System}
\textcolor{black}{As anticipated in Sec.\ref{sec:bg}}, in the case of the linear systems we are able to get an explicit expression for entropy production rate because an explicit expression of the free propagator $\cW^{(0)}_t(X|Y)$ is available, i.e.
\begin{eqnarray}
&\cW^{(0)}_t(X|Y) = \cG_{\cC_t}(X - e^{-t A} Y) \\
&\cC_t= \int_0^t dt'\, e^{-t'A} \Sigma e^{-t'A^T} \qquad \overset{t\to\infty}{\longrightarrow} \qquad \cC =  \int_0^\infty dt\, e^{-tA} \Sigma e^{-tA^T}
\end{eqnarray}
from which it \textcolor{black}{is} easy to get the propagator and its reverse in presence of the jumps
\begin{flalign}
    \cW^{(n)}_t(X|Y,\cK_n) &= \int \prod_{k=1}^n \lx[dX_k \cG_{\cC_{t_k-t_{k-1}}}(X_k - U_k - e^{-(t_k-t_{k-1})A}X_{k-1}) \rx] \cG_{\cC_{t-t_n}}(X-e^{-(t-t_n)A}X_n) = \nonumber\\
    &= \cG_{\cC_t}(X-e^{-tA}Y - \sum_{k=1}^n e^{-(t-t_k)A} U_k) \\ 
    \overleftarrow{\cW}^{(n)}_t(Y|X,\cK_n) &= \int \prod_{k=1}^n \lx[dX_k \cG_{\cC_{t_k-t_{k-1}}}(X_{k-1} - e^{-(t_k-t_{k-1})A}(X_k - U_k) \rx] \cG_{\cC_{t-t_n}}(X_n-e^{-(t-t_n)A}X) = \nonumber\\
    &= \cG_{\cC_t}(Y-e^{-tA}X + \sum_{k=1}^n e^{-t_kA} U_k)
\end{flalign}
This implies that, given the jumps and the starting point $Y$, the entropy production rate is
\begin{equation}
    \cS^{(n)}_t(Y,\cK_n) = \frac{1}{2t} \int dZ\, \cG_{\cC_t}(Z) W(Z)^T \cC_t^{-1} W(Z) - \frac{N}{2t} \\
\end{equation}
where
\begin{eqnarray}
W(Z)&= Y - e^{-t A} X  + \sum_{k=1}^n e^{-t_k A} U_k = \nonumber \\
&= -e^{-tA}\lx(Z -(e^{t A} + e^{-tA})Y - \sum_{k=1}^n \lx(e^{(t-t_k)A}-e^{-(t-t_k)A}\rx)U_k \rx)
\end{eqnarray}
Now, we average over the jumps by using $\overline{U_j U_k^T}=\Gamma\delta_{j,k}$ and by observing that, from equation \eqref{avejumpsum} we have
\begin{flalign}
\overline{\sum_{k=1}^n \lx(\cM(t-t_k)\cL(t)\cN(t-t_k)\rx)_{ij}} &=
\int d\cK_n \cP(\cK_n) \sum_{k=1}^n \lx(\cM(t-t_k)\cL(t)\cN(t-t_k)\rx)_{ij} =\nonumber\\
&= \sum_{lm}\cL_{lm}(t) \overline{\sum_{k=1}^n \cM_{il}(t-t_k)\cL_{lm}(t) \cN_{mj}(t-t_k)} \nonumber\\
&= \lambda \sum_{lm} \cL_{lm}(t)\int_0^t dt' \cM_{il}(t')\cN_{mj}(t') \nonumber\\
&= \lambda \lx(\int_0^t dt'\, \cM(t')\cL(t)\cN(t')\rx)_{ij}
\end{flalign}
then, to eliminate the dependence on the initial state $Y$ we can compute the covariance matrix $\widehat{\cC} = \overline{\ave{XX^T}}$ over the stationary measure, i.e.
\begin{eqnarray}
    &\overline{\ave{X|Y}_t} = e^{-tA} Y \quad \overset{t\to\infty}{\longrightarrow} \quad \overline{\ave{X}} = 0 \\
    &\widehat{\cC}_t =\overline{\ave{XX^T|Y}_t - \ave{X|Y}_t \ave{X|Y}_t^T} = \int_0^t dt' \, e^{-t'A} \lx(\Sigma + \lambda \Gamma \rx) e^{-t' A^T} \\
    &\widehat{\cC}_t \quad \overset{t\to\infty}{\longrightarrow} \quad \widehat{\cC} = \overline{\ave{XX^T}} = \int_0^\infty dt \, e^{-tA} \lx(\Sigma + \lambda \Gamma \rx) e^{-t A^T} = \cC + \lambda \cC'\\
    &\cC' =  \int_0^\infty dt \, e^{-tA} \Gamma e^{-t A^T}
\end{eqnarray}
So, by putting it all together we get
\begin{eqnarray}
    \cS_t &=& \text{Tr}\lx[\cC_t e^{-tA^T} \cC_t^{-1} e^{-tA} + \widehat{\cC} (1-e^{-2tA^T})\cC_t^{-1} (1-e^{-2tA}) - I\rx]/2t+ \nonumber\\
    &+& \text{Tr}\lx[\Gamma \int_0^t dt' \lx(e^{t'A^T}-e^{-t'A^T}\rx) e^{-t A^T}\cC_t^{-1}e^{-t A}\lx(e^{t' A}-e^{-t' A}\rx)\rx] /2t\,.
\end{eqnarray}
Finally, the limit $t \to 0$ is made by considering $\cC_t \simeq \Sigma t$
and $e^{-tA}\simeq 1 - tA$ which leads to
\begin{equation}
    \cS = \text{Tr}\lx[2 \widehat{\cC}A^T \Sigma^{-1}A - A \rx] = 
     \text{Tr}\lx[2 \cC A^T \Sigma^{-1}A - A \rx] + 2\lambda \text{Tr} \lx[ \cC' A^T \Sigma^{-1} A \rx]\,.
\end{equation}
It is easy to prove that, the Onsager's equilibrium condition $A\widehat{\cC}=\widehat{\cC}A^T$ still leads to a positive entropy production rate.
In fact, since $A\widehat{\cC}+ \widehat{\cC}A^T=2\widehat{\cC}A^T=\Sigma+\lambda \Gamma$ we have
\begin{equation}
\cS = \lambda \text{Tr}\lx[\Gamma \Sigma^{-1}A\rx]\,.
\end{equation}
We note that this value correspond to the minimum of the entropy production. Indeed, 
\begin{align}
    &2\widehat{\cC}A^T=\Sigma+\lambda \Gamma+\Delta\,\\
    &\Delta = \widehat{\cC}A^T-A\widehat{\cC}
\end{align}
lead to
\begin{equation}
\cS = \cS(\Delta)= \lambda \text{Tr}\lx[\Gamma \Sigma^{-1}A\rx] + \text{Tr}\lx[\Delta \Sigma^{-1}A\rx]
\end{equation}
The variation of $\cS$ due to a change in $\Delta$ can be written as

\begin{equation}
\delta \cS = \text{Tr}\lx[(\delta\Delta) \Sigma^{-1}A\rx]=\sum_{ijk}(\delta\Delta)_{ij}\Sigma_{jk}^{-1}A_{ki}=\sum_{i<j}(\delta\Delta)_{ij}\lx[\lx(\Sigma^{-1}A\rx)_{ji}-\lx(\Sigma^{-1}A\rx)_{ij}\rx]
\end{equation}
In cases where $A\Sigma\neq\Sigma A^T$, $\delta \cS$ can not be equal to $0$ and the minimum is obtained for $\Delta =0$, that is $\widehat{\cC}A^T=A\widehat{\cC}$. Interestingly we have $\delta \cS=0$ whenever $A\Sigma=\Sigma A^T$. Note that the last condition does not imply Onsager relation. In fact,
\begin{align}
&A\widehat{\cC}=\int_0^\infty dt \, e^{-tA} A\lx(\Sigma+\lambda\Gamma\rx) e^{-t A^T}\,,\\
&\widehat{\cC}A^T=\int_0^\infty dt \, e^{-tA} \lx(\Sigma+\lambda\Gamma\rx)A^T e^{-t A^T}=\lambda\Gamma-\int_0^\infty dt \, e^{-tA} A\lambda\Gamma e^{-t A^T}+\int_0^\infty dt \, e^{-tA} A\Sigma e^{-t A^T}
\end{align}
where the last equation has been obtained using integration by parts and imposing $A\Sigma=A^T\Sigma$. Thus, $\Delta$ takes the form
\begin{align}
\Delta&= \lambda\Gamma-2 \int_0^\infty dt \, e^{-tA} A\lambda\Gamma e^{-t A^T}=2 \int_0^\infty dt \, e^{-tA} \lambda\Gamma A^T e^{-t A^T}-\lambda \Gamma =\nonumber\\
&=\lambda \int_0^\infty dt \, e^{-tA} \lx(\Gamma A^T- A\Gamma\rx)e^{-t A^T}
\end{align}
which is not necessarily identical to the null operator.
\section{Gradient Systems}\label{app:Gradient-Systems}
The procedure described in~\ref{appendix:EPR} is rigorous but analytical computations can rarely be carried out. Nevertheless, it is possible to obtain expressions suitable for numerical computations for gradient systems in contact with a single thermal bath. \textcolor{black}{These expressions are those provided in Sec.\ref{sec:ep} for the one-dimensional linear case as well as for the particle moving on periodic potentials}. Consider a process $X$ whose dynamic is 
\begin{equation}
    \dot{X} = -\partial_X V(X) + \sqrt{2T}\xi(t) +\sum_j U_j \delta(t-t_j)\\
\end{equation}
with $\ave{\xi(t)\xi(t')^T}=I\delta(t-t')$. Between two jumps, the entropy production is 
\begin{equation}
    S_t(X,Y) = \frac{V(Y) -V(X)}{T}.\\
\end{equation}
The entropy production rate once $t$ and $\cK_n$ are given is 
\begin{equation}
\cS^{(n)}_t(X_0,X_t,\cK_n)=  \frac{1}{tT}\sum_{k=1}^n \lx(V(X_k+U_k)-V(X_k)\rx)+\frac{V(X_0) -V(X_t)}{tT}
\label{eq:appendix_gradient}
\end{equation}
where the last term vanishes when $t\to+\infty$.
The average entropy production rate is
\begin{equation}
\cS = \frac{1}{tT}\overline{\ave{\sum_{k=1}^n \lx(V(X_k+U_k)-V(X_k)\rx)}} = \frac{1}{tT} \sum_{n=0}^\infty \cP_\lambda(n|t) \sum_{k=1}^n \sum_{l=1}^\infty \frac{\ave{V^{(l)}(X_k)}}{l!}\overline{U_k^{l}}
\end{equation}
with
\begin{equation}
\sum_{l=1}^\infty \frac{V^{(l)}(X)}{l!}U^{l}=\sum_{l=1}^{\infty}\sum_{\{l_k\}:\sum l_k =l}\frac{\partial^{(l)}V(X)}{\partial_{x_1}^{(l_1)}\cdots\partial_{x_n}^{(l_n)}}\frac{U_{x_1}^{l_1}\cdots U_{x_n}^{l_n}}{l_1!\cdots l_n!}
\end{equation}
\textcolor{black}{As mentioned in Sec.~\ref{sec:ep}, it is also possible to include a constant pulling force $f$ in the potential, i.e. $V_f(X)=V(X)-f\cdot X$. This modification only affects the boundary term on the right hand side of Eq.~\ref{eq:appendix_gradient} where it appears a term proportional to $f\cdot (X_0-X_t)$. On average, this term converges to $(f\cdot j_s) t$ giving rise to the first term on the right hand side of Eq.~\ref{eq:relation_epr_current}.}
\subsection{One dimensional Systems}
For one dimensional systems it is possible to further simplify the expression for $\cS$. By using the explicit formula for Gaussian moments
\begin{equation}
\overline{U^{2m}} = \frac{\lx(2m\rx)!}{m!}\lx(\frac{\sigma^2}{2}\rx)^m \qquad \overline{U^{2m+1}} = 0 \\
\end{equation}
one gets
\begin{align}
\cS &= \frac{1}{tT}\overline{\ave{\sum_{k=1}^n \lx(V(X_k+U_k)-V(X_k)\rx)}} = \frac{1}{tT} \sum_{n=0}^\infty \cP_\lambda(n|t) n \sum_{l=1}^\infty \frac{\ave{V^{(l)}(X)}}{l!}\overline{U^{l}} = \nonumber \\
&=\frac{\lambda}{T}\sum_{m=1}^\infty \frac{\ave{V^{(2m)}(X)}}{m!}\lx(\frac{\sigma^2}{2}\rx)^m
\end{align}
This formula can not be simplified anymore without specifying the potential $V(X)$. For quadratic, periodic or quartic potentials one has
\begin{eqnarray}
    V(x)=\frac{1}{2}\eta x^2 &\rightarrow& \eta\frac{\lambda\sigma^2}{2T}\\
    V(x)=1-\cos{\frac{2\pi}{L}x} &\rightarrow& \cS = \frac{\lambda}{T}\overline{\ave{\cos{\frac{2\pi}{L}x}}} \lx(1-e^{-2\lx(\pi \sigma/L\rx)^2}\rx) \\
    V(x)= \frac{\alpha}{4} x^4-\frac{\beta}{2} x^2 &\rightarrow& \cS = 3\alpha\frac{\lambda\sigma^2}{2T} \lx(\overline{\ave{x^2}}+\frac{\sigma^2}{2}-\frac{\beta}{3\alpha} \rx)\,.
\end{eqnarray}
\section*{References}
\bibliographystyle{unsrt.bst}
\bibliography{biblio}
\end{document}